\documentclass[12pt]{article}
\usepackage[utf8]{inputenc}
\usepackage[T1]{fontenc} 
\usepackage{mathptmx} 
\usepackage[autostyle=true]{csquotes} 

\usepackage{natbib}

\usepackage{amsmath,amsfonts,amssymb}
\usepackage{graphicx}%
\usepackage{color}
\usepackage[dvipsnames]{xcolor}
\usepackage{ulem}
\usepackage{enumerate}
\usepackage{url}
\usepackage{bm}
\usepackage{bbm}
\usepackage{float}
\usepackage{mathtools}
\usepackage[toc,page]{appendix}
\usepackage{setspace}
\usepackage{subcaption}
\usepackage{lipsum}
\usepackage{needspace}
\usepackage{xpatch}
\usepackage{gensymb}
\usepackage{cancel}
\usepackage{siunitx}
\usepackage{hyperref}
\usepackage{authblk}

\usepackage{geometry}
\xpretocmd{\section}{\needspace{6\baselineskip}}{}{}
\usepackage[section]{placeins}

\newread\tmp

\newcommand{\Sec}[2]{\section{#1\label{Sec:#2}}}
\newcommand{\sSec}[2]{\subsection{#1 \label{Sec:#2}}}
\newcommand{\ssSec}[2]{\subsubsection{#1 \label{Sec:#2}}}

\newcommand{\setR}{\mathbbm{R}}

\newcommand{\conv}[1]{\stackrel{#1}{\ast}}
\newcommand{\pare}[1]{\left(\, #1 \, \right)}

\newcommand{\bra}[1]{\left[\, #1 \, \right]}
\newcommand{\abs}[1]{\left|\, #1 \, \right|}

\newcommand{\Set}[1]{\left\{\, #1 \, \right\}}

\newcommand{\vect}[1]{\pare{\begin{array}{ccc}#1\end{array}}}
\newcommand{\ve}{\vec{e}}

\newcommand{\vcF}{\vec{\cF}}

\newcommand{\vcC}{\vec{\cC}}
\newcommand{\vcR}{\vec{\cR}}
\newcommand{\vcX}{\vec{\cX}}

\newcommand{\vnabla}{\vec{\nabla}}
\newcommand{\diag}{\mathop{\mathrm{diag}}}

\newcommand{\CRep}[2]{\varpi_{#1#2}}

\newcommand{\Cell}[2]{{#1}_{#2}} 
\newcommand{\V}[2]{V_{\Cell{#1}{#2}}} 
\newcommand{\K}[2]{\cK_{\Cell{#1}{#2}}} 
\newcommand{\M}[4]{M^{\Cell{#1}{#2}}_{\Cell{#3}{#4}}}

\newcommand{\E}[3]{\cE^{(#1)}_{\Cell{#2}{#3}}}
\newcommand{\Vdr}[3]{\cV^{(#1)}_{\Cell{#2}{#3}}}
\newcommand{\F}[2]{F_{\Cell{#1}{#2}}}
\newcommand{\z}[2]{\zeta_{\Cell{#1}{#2}}}

\newcommand{\N}[1]{\cN^{(#1)}}

\newcommand{\Co}[4]{\Gamma^{\Cell{#1}{#2}}_{\Cell{#3}{#4}}}

\newcommand{\W}[4]{W^{\Cell{#1}{#2}}_{\Cell{#3}{#4}}}

\renewcommand{\t}[2]{\tau^{#1}_{#2}}

\newcommand{\Amp}[2]{w^{#1}_{#2}}

\newcommand{\D}[4]{d\bra{\Cell{#1}{#2}, \, \Cell{#3}{#4}}}
\newcommand{\Dd}[4]{d^2\bra{\Cell{#1}{#2}, \, \Cell{#3}{#4}}}

\newcommand{\cC}{{\mathcal C}}
\newcommand{\cD}{{\mathcal D}}
\newcommand{\cE}{{\mathcal E}}

\newcommand{\cK}{{\mathcal K}}

\newcommand{\cI}{{\mathcal I}}
\newcommand{\cL}{{\mathcal L}}

\newcommand{\cM}{{\mathcal M}}
\newcommand{\cN}{{\mathcal N}}
\newcommand{\cP}{{\mathcal P}}

\newcommand{\cR}{{\mathcal R}}
\newcommand{\cS}{{\mathcal S}}
\newcommand{\cF}{{\mathcal F}}
\newcommand{\cX}{{\mathcal X}}

\newcommand{\cV}{{\mathcal V}}
\newcommand{\cT}{{\mathcal T}}
\newcommand{\cTm}{{\mathcal T}^{-1}}
\newcommand{\cW}{{\mathcal W}}

\newcommand{\veta}{\vec{\eta}}
\newcommand{\vz}{\vec{0}}
\newcommand{\vun}{\vec{1}_N}

\newcommand{\vcP}{\vec{\cP}}

\newcommand{\vk}{\vec{k}}
\newcommand{\vV}{\vec{V}}

\newcommand{\vphi}{\vec{\phi}}

\newcommand{\vpsi}{\vec{\psi}}

\title{How does the inner retinal network shape the ganglion cells receptive field : a computational study}

\author[1,2]{\normalsize Evgenia Kartsaki}
\author[2,3]{\normalsize Gerrit Hilgen}
\author[2]{\normalsize Evelyne Sernagor}
\author[1]{\normalsize Bruno Cessac}

\affil[1]{\small Université Côte d'Azur, Inria, Biovision team and Neuromod Institute, 
Sophia Antipolis, France}
\affil[2]{\small Biosciences Institute, Newcastle University, Newcastle upon Tyne NE2 4HH, UK}
\affil[3]{\small Health and Life Sciences, Applied Sciences, Northumbria University, Newcastle upon Tyne NE1 8ST, UK}

\date{}

\begin{document}

\maketitle


\begin{abstract}

We consider a model of basic inner retinal connectivity where bipolar and amacrine cells interconnect, and both cell types project onto ganglion cells, modulating their response output to the brain visual areas.   
We derive an analytical formula for the spatio-temporal response of retinal ganglion cells to stimuli taking into account the effects of amacrine cells inhibition. 
This analysis reveals two important functional parameters of the network: (i) the intensity of the interactions between bipolar and amacrine cells, and, (ii) the characteristic time scale of these responses. Both parameters have a profound combined impact on the spatiotemporal features of RGC responses to light. 
The validity of the model is confirmed by faithfully reproducing pharmacogenetic experimental results obtained by stimulating excitatory DREADDs (Designer Receptors Exclusively Activated by Designer Drugs) expressed on ganglion cells and amacrine cells subclasses, thereby modifying the inner retinal network activity to visual stimuli in a complex, entangled manner. Our mathematical model allows us to explore and decipher these complex effects in a manner that would not be feasible experimentally and provides novel insights in retinal dynamics.
\end{abstract}


\newpage

\Sec{Introduction}{Introduction}

Visual processing involves some of the most complex neural networks in the vertebrate central nervous system  \citep{marr:82,chalupa-werner:04,besharse-bok:11,daw:12}.
The retina is the  entry point to our visual system. Located at the back of the eye, this thin neural tissue receives the light that the cornea and lens have captured from different parts of the visual scene, converts it into electrical signals and finally, transmits these signals to the brain visual areas. In particular, light follows a vertical excitatory pathway in the retina, from photoreceptors (PRs) to bipolar cells (BCs) and onwards to retinal ganglion cells (RGCs), modulated laterally by inhibitory interneurons; horizontal (HCs) and amacrine cells (ACs) (Figure \ref{Fig:Connectivity}, left). RGCs serve as a bridge between the retina and the brain, conveying highly processed and integrated signals from the upstream retinal neurons to downstream visual processing cortical areas. Amazingly, the human brain can recreate images from interpreting parallel streams of information of about one million RGCs, the sole retinal output neurons. This ability is partially due to the astonishing functional, anatomical and molecular diversity across the retinal layers. However, how the different cell classes interact to ultimately encode, via RGCs, a visual scene into spike trains (action potentials) deciphered by the brain remains largely a mystery. 

RGCs are indeed embedded in a complex network and their response is roughly driven by two "controllers": 1) The output of BCs that includes both the intrinsic response properties of these cells and the actions of HCs and ACs upon them (lateral connectivity); 2) The direct input from ACs via chemical synapses or gap junctions, helping spike synchrony between neighbor RGCs \citep{Masland2012, Demb2015b}.
As a consequence, the response of RGCs to visual stimuli does not only depend on the physiological characteristics of the cells \citep{Sanes2015}, but also on the network they are embedded in, and, on the stimulus itself \citep{cessac:22}. 
Previous studies have thus attempted to investigate how the inner retinal neurons are organised into parallel circuits across different cell types and converge onto RGCs \citep{Wassle2004, Gollisch2010}. This has been studied extensively at the level of bipolar cells, leading to a fairly good understanding of their function \citep{Euler2014}. Other studies have investigated the functional role of amacrine cell (ACs) types in retinal processing \citep{Asari2012, Franke2017a, Diamond2017,schroder-klindt-etal:20}, suggesting either specific functions such as direction selectivity (starbust ACs) or more general computations, like motion anticipation \citep{berry-brivanlou-etal:99, souihel-cessac:21}.
A number of studies have also focused on how retinal interneurons contribute towards the sensitivity to activity patterns of RGCs. For example, \cite{deVries2011} studied the spatio-temporal effects of ACs on RGCs outputs and proposed mechanisms related to predictive coding of the retina. \cite{Manu2022}, \cite{Ichinose2014},  manipulated HCs and ACs, the first by injecting patterns of current and the latter by using pharmacology, to directly measure the effect on retinal receptive fields surround. Likewise, \cite{protti-di-marco-etal:14} studied the spatial organisation of excitatory and inhibitory synaptic inputs onto ON and OFF RGCs in the primate retina using pharmacology. 
Nevertheless, the potential role of the BCs-ACs network on the RGCs response, both from a theoretical and experimental perspective, hasn't been sufficiently explored yet. The scope of the present study is to make one step further in this direction.

Numerous models of the retina have been proposed, with different levels of biological detail and across multiple spatial and temporal scales. Phenomenological models have become quite popular due to their simplicity and efficiency to fit to experimental data and explain observations, such as the linear-nonlinear model \citep{Chichilnisky2001, Paninski2003, Simoncelli2004, Schwartz2006}, the generalized linear models \citep{Pillow2008} or black-box convolutional neural networks of retinal function \citep{McIntosh2017}. However, these models are agnostic to the biophysical details that underlie the input-output transform of the visual system. On the other hand, mechanistic models exhibit a direct relationship between the underlying biophysics of neurons and the parameters of the model. Well known examples are the Hudgkin-Huxley \citep{Hodgkin1952}, the Fitzhugh-Nagumo \citep{fitzhugh:69,nagumo-arimoto-etal:62}, the Morris-Lecar \citep{morris-lecar:81} models but they were mainly designed to describe the spike generation while most of the neurons in the retina (except RGCs) are not spiking.  In addition, this type of models turns out to be cumbersome to analyze and simulate, especially when dealing with collective network dynamics.

The model developed in the present work falls into both categories and it is definitely inspired by some previous work  \citep{berry-brivanlou-etal:99,Chen2013, souihel-cessac:21}. It is a multi-stage, phenomenological model, with simplifications regarding the complex retinal structure. Nevertheless, it aims to be relatively precise from a biological perspective, by realistically reproducing RGCs’ responses to light from the experimental recordings. Our primary goal is to gain insights about the underlying biophysical processes giving rise to certain experimentally observed phenomena, rather than proposing an exact model of the retina.

On a theoretical ground, we consider a network of BCs connected via ACs, both cell types being connected with chemical synapses to RGCs. Based on a mathematical study, we derive an analytical formula of the RGCs receptive field (RF) that takes into account the lateral ACs connectivity and shows how the response of RGCs to spatio-temporal stimuli is shaped. Especially, we emphasize the role of: (i), the average intensities of the interactions BCs - ACs and ACs-BCs, and, (ii), the characteristic time scales of these cells response. 
Varying these parameters acts on the shape of the response to light with potential prominent effects such as a switch from monophasic to biphasic in the temporal RF.

We illustrate our predictions by analysing experimental data obtained from the pharmacological action of excitatory DREADDs (Designer Receptors Exclusively Activated by Designer Drugs) in genetically modified mice. DREADDs are activated by "designer drugs" such as clozapine-n-oxide (CNO), resulting in an increase in free cytoplasmic calcium and profound increase in excitability in the cells that express these DREADDs \citep{Roth2016}. We found DREADD expression both in subsets of RGCs and in ACs in the inner nuclear layer (INL) \citep{hilgen-kartsaki-etal:22}. In these conditions, CNO would act both on ACs and RGCs providing a way to unravel the entangled effects of: (i) direct excitation of DREADD-expressing RGCs and increase inhibitory input onto RGCs originating from DREADD-expressing ACs; (ii) change in cells response time scale (via a change in the membrane conductance), thereby providing an experimental set up to validate our theoretical predictions.

In the following, we propose a simplified model for the BCs - ACs - RGCs network. The concerted activity of BCs - ACs -RGCs in response to a visual stimulus is described by a large dimensional dynamical system whose mathematical study allows an explicit computation of the RF of BCs, ACs, RGCs and, more generally, to anticipate the effects resulting from light stimulation conjugated with the network activity due to ACs lateral inhibition, in control conditions and in the presence of CNO.  Computing the receptive fields of all cell types (especially RGCs) allows us to to disentangle the concerted effect of  ACs lateral connectivity and CNO on the RF   of RGCs and provides an excellent agreement with experimental data. 
We argue, on the basis of the model and analytical computations, that the BCs-ACs network shapes the RF of RGCs via two main reduced dimensionless parameters, one characterizing the ratio in the time scales of ACs and BCs response, and the other, characterizing the ratio between the synaptic weight BCs $\to$ ACs and ACs $\to$ BCs. This defines a two dimensional map, called the "RFs map". We show how the experimental data settle in this map and we link the observed changes in the shape of the RF, when applying CNO, to a displacement in this map. Finally, we extrapolate our model predictions to situations  where the RGCs response could not be explained by the physiological properties of the cell but would involve the ACs network.

\section{Methods}\label{Sec:Methods}


\sSec{The retina model}{RetinaModel}

\ssSec{Structure}{Structure}

\begin{figure}
\centering
\includegraphics[height=10cm,width=12cm]{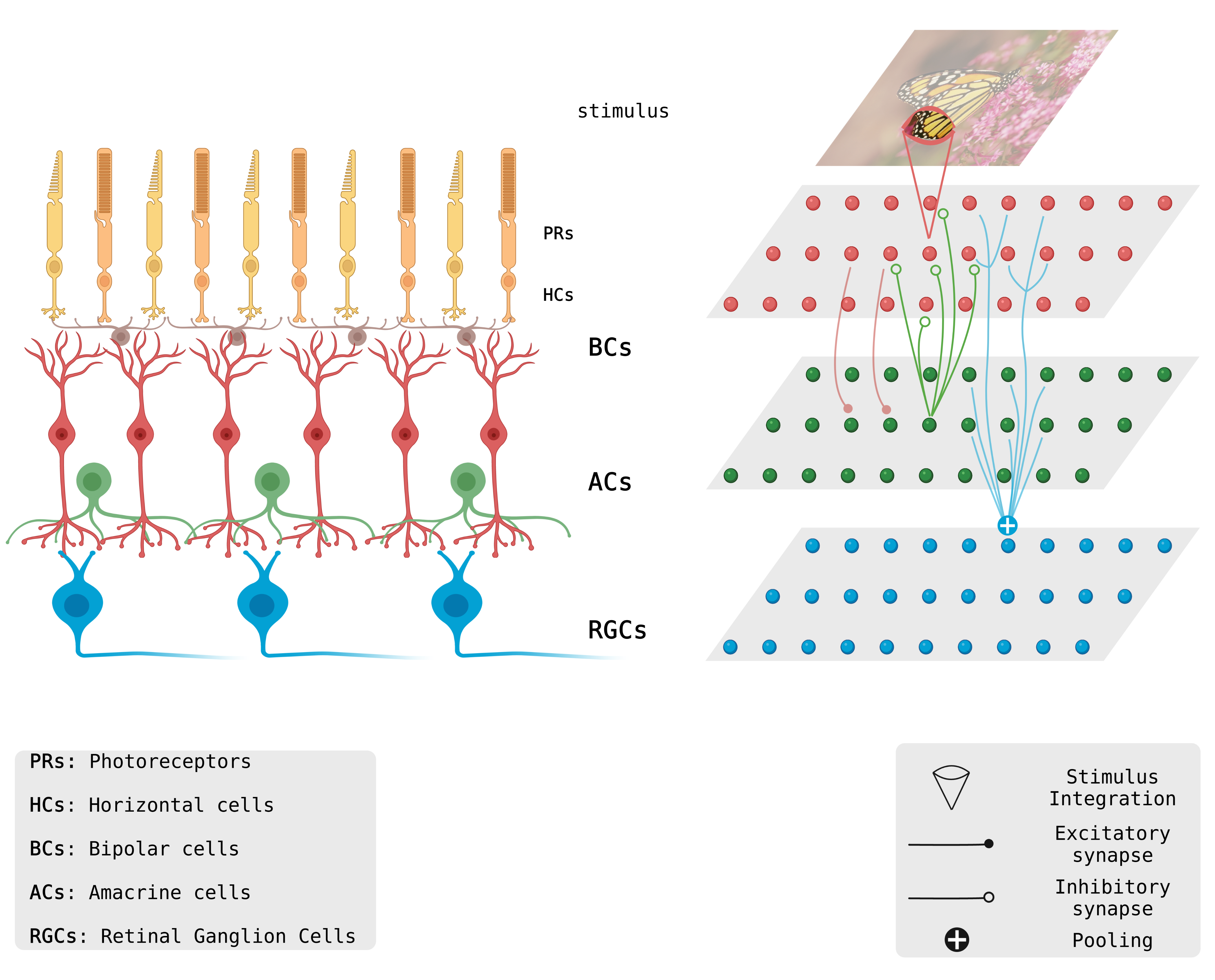}
\caption{\small\doublespacing \textbf{Translation of the retinal circuit to a computational network model.} \textbf{Left.} Schematic of the retina. Light activates the photoreceptor cells (PRs), that transduce the input into a cascade of biochemical and electrical events that can stimulate BCs and onwards RGCs. This vertical excitatory pathway is modulated by inhibitory interneurons comprising two groups; horizontal (HCs) and amacrine cells (ACs). All these neural signals are integrated by RGCs and finally converted into action potentials going to the brain. \textbf{Right}. Schematic view of the model. The joint integration of PRs and HCs in a limited region of space is modelled by a spatio-temporal kernel mathematically defining the OPL input to a BC, corresponding to eq. \eqref{eq:Vdrive}. The convolution of this kernel with the stimulus drives the BC evolution, modulated by inhibitory connections with ACs. BCs indeed make excitatory synaptic connections with ACs and ACs inhibit BCs. Finally, RGCs pool over many BCs and ACs in their neighbourhood. Note that the cells
connections on the right panel are just shown for illustration and do not necessarily correspond to the connectivity used in the model. Cells are organised along a square lattice viewed here in a 3D perspective.
}
\label{Fig:Connectivity}
\end{figure}

We assimilate the retina to a superimposition of $3$ layers, each one corresponding to a cell type (BCs, ACs, RGCs), and
being a flat, two-dimensional square of edge length $L$ mm where spatial coordinates are noted $x,y$ (Figure ~\ref{Fig:Connectivity}). 

We consider a simplified form of connectivity, inspired from the real connectivity in the retina, illustrated in Figure \ref{Fig:Connectivity}.  First, there are as many BCs as ACs and RGCs ($N$ cells per type so that the total number of cells is $3N$). 
BCs are labelled with an index $i=1 \dots N$, ACs with an index $j=1 \dots N$, RGCs with an index $k=1 \dots N$.  BCs are connected to ACs. We note $\W{A}{j}{B}{i} \leq 0$ the weight of the  inhibitory synapse from AC $j$ to BC $i$. We use the convention that $\W{A}{j}{B}{i}=0$ when AC $j$ is not connected to BC $i$. Likewise, we note $\W{B}{i}{A}{j} \geq 0$ the weight of the excitatory synapse from BC $i$ to AC $j$,  $\W{B}{i}{G}{k} \geq 0$ the weight of the excitatory synapse from BC $i$ to RGC $k$ and $\W{A}{j}{G}{k} \leq 0$ the weight of the  inhibitory synapse from AC $j$ to RGC $k$. We note $\W{A}{}{B}{}$ the (square) matrix of connections from ACs to BCs and so on. Note that the model affords as well connections between the same cell types (e.g. ACs to ACs or RGCs to RGCs via gap junctions) but it would consequently complicate the theoretical analysis.

\ssSec{Visual input}{Input}

Each neuron  has a receptive field (RF), a specific region of the visual field where light stimulus will modify the activity of this neuron. The term RF is not limited only to the spatial region but it is often extended to include the temporal structure within this region. The RF usually exhibits a centre-surround organisation shared by many cell types in the retina \citep{Masland2012a}. Each BC receives synaptic inputs from its upstream circuitry, a combination of dendritic excitatory inputs from photoreceptors (rods and cones) and inhibitory inputs from horizontal cells \citep{Franke2017a} occurring at the level of the Outer Plexiform Layer (OPL) and this interaction emerges on their RF. 

A popular approach to simplify the complex process involved is to model the RF as a single
spatio-temporal linear filter that essentially represents the opposition between the centre of the receptive field, driven by photoreceptors, and the surround signal transmitted by horizontal cells. The membrane potential of the BC’s soma can then be linearly approximated by the convolution of the spatio-temporal kernel $\K{B}{i}$, featuring the biophysical processes at the OPL, with the visual stimulus $\cS(x,y,t)$. As we do not consider color sensitivity here, $\cS$ characterizes a black and white scene, with a control on the level of contrast $\in [0,1]$.  The voltage of BC $i$ is stimulus-driven by the term ("OPL input"):

\begin{equation}\label{eq:Vdrive}
\begin{split}
\Vdr{drive}{i}{}(t) &= \bra{\K{B}{i} \conv{x,y,t} \cS}(t) \\
&= \int_{x=-\infty}^{+\infty} \,\int_{y=-\infty}^{+\infty} \,
\int_{s=-\infty}^{t} \,  \cK(x-x_i,y-y_i,t-s) \, \cS(x,y,s) \, dx \, dy \, ds,
\end{split} 
\end{equation}
where $\conv{x,y,t}$ means space-time convolution. For simplicity we consider only one family of BCs so that the kernel $\cK$ is the same for all BCs. What changes is the center of the RF, located at $x_i,y_i$, which also corresponds to the coordinates of the BC $i$.   
We consider in the paper a separable kernel
$\cK(x,y,t)=\cK_S(x,y) \, \cK_T(t)$ where $\cK_S$ is the spatial part and $\cK_T$ the temporal part. We restrict to ON or OFF BCs with a \textit{monophasic} spatio-temporal kernel of the form:
\begin{equation}\label{eq:KT}
\K{}{T}=\pare{A_0 \, \frac{t^2}{2 \, \tau_{RF}^{3}} e^{-\frac{t}{\tau_{RF}}} \, + \, b_0} \, H(t).
\end{equation}
where $H(t)$ is the Heaviside function. The parameter $A_0$ controls the amplitude of the OPL input and can be positive (ON BC) or negative (OFF BC), while $b_0$ controls the level of residual polarisation observed in experiments, when light stimulation has stopped. This parameter is further explained below.  Experimentally, the BC temporal response can be biphasic in time, because of
receptor desensitization and because HCs feedback on to
PRs terminals. Note however, that the BCs response can also be monophasic \citep{Shreyer2018}. As we precisely want to point out the effect of lateral inhibition on the BCs response, we consider a monophasic temporal OPL input and show that, tuning the lateral connectivity parameters in the model, one can indeed obtain biphasic shapes, as expected, but also more complex dynamical responses.

\begin{figure}
\centerline{
\includegraphics[width=0.5\textwidth,height=0.3\textheight]{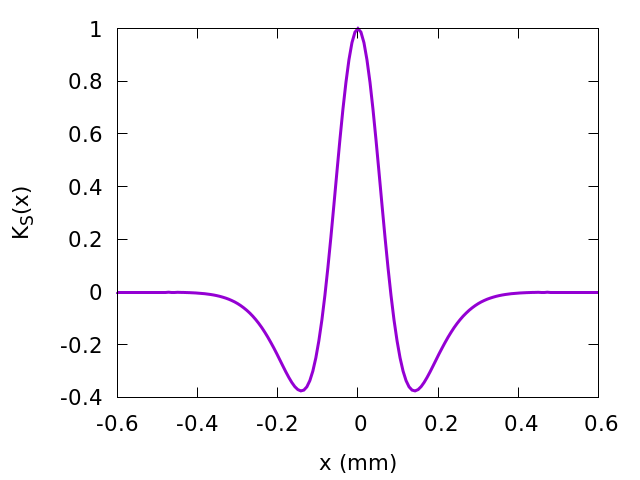}
\hspace{1cm}
\includegraphics[width=0.5\textwidth,height=0.3\textheight]{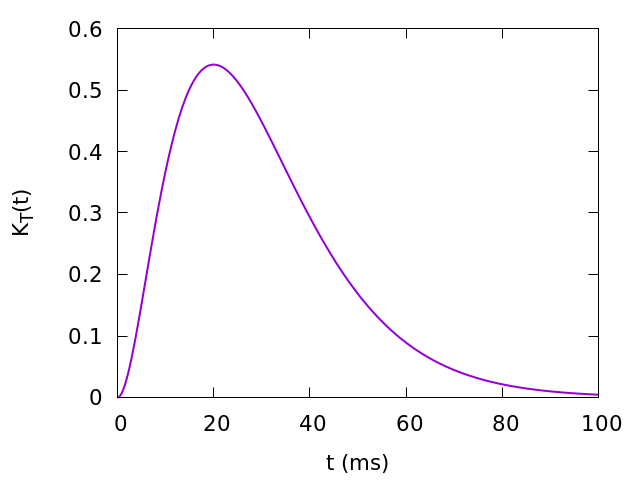}
}

\caption{\doublespacing\textbf{OPL kernel.} \textbf{Left.} Spatial part in one spatial dimension. \textbf{Right.} Temporal part.}
\label{Fig:OPL_Kernel}
\end{figure}

The spatial part, $\K{}{S}$, is a classical difference of Gaussians.
This spatio temporal kernel is illustrated in Figure  \ref{Fig:OPL_Kernel}.

\ssSec{Voltage dynamics}{VoltageDyn}

In the model, neurons are characterized by their voltage. Although RGCs are spiking, we will be indeed interested only in their voltage variations, as a function of the stimulus and network effects. Spiking could be obtained from voltage e.g. using  a Linear Nonlinear Poisson process (LNP) for spiking neurons \citep{berry-brivanlou-etal:99,chen-marre-etal:13,souihel-cessac:21}. We note $\V{B}{i}$ the voltage of BC $i$, $\V{A}{j}$ the voltage of AC $j$, $\V{G}{k}$ the voltage of RGC $k$. The joint dynamics of voltages is given by the dynamical system:
\begin{equation}\label{eq:Diff_Syst}
\left\{
	\begin{array}{lll}
\frac{d\V{B}{i}}{d t} &=& - \frac{1}{\tau_{B_i}} \V{B}{i} + \sum_{j=1}^{N_A} \W{A}{j}{B}{i}\,\N{A}\pare{\V{A}{j}} +  \F{B}{i}(t), \quad i=1 \dots N\\
	&&\\
\frac{d\V{A}{j}}{d t} &=& - \frac{1}{\tau_{A_j}} \V{A}{j} + \sum_{i=1}^{N_B} \W{B}{i}{A}{j}\,\N{B}\pare{\V{B}{i}}  + \z{A}{}, \quad j=1 \dots N,\\
&&\\
\frac{d \V{G}{k}}{d t} 
&=& - \frac{1}{\tau_{G_k}} \V{G}{k}+ \sum_{i=1}^{N_B} \W{B}{i}{G}{k}  \cN_B(\V{B}{i}) + \sum_{j=1}^{N_A} \W{A}{j}{G}{k}  \cN_A(\V{A}{j}) + \z{G}{}, \quad k=1 \dots N.
	\end{array}
	\right.
\end{equation}
This form is derived, from first principles and using simplifying hypotheses, in the section \ref{Sec:Generic} of the supplementary material.
%
The rectification term:
\begin{equation}\label{eq:Rectif}
\cN_A(\V{A}{j}) = \left\{\begin{array}{lll}
\V{A}{j}-\theta_A, \quad &\mbox{if} \,\, \V{A}{j} > \theta_A;\\
0, \quad &\mbox{otherwise}  
\end{array} \right.,
\end{equation}
ensures that the synapse $j \to i$ is inactive when the pre-synaptic voltage $\V{A}{j}$ is smaller than the threshold  $\theta_A$. The same holds for the rectification term $\cN_B(\V{B}{i})$ with a threshold $\theta_B$.

The term:
\begin{equation}\label{eq:FBip}
 \F{B}{i}(t)= \frac{\Vdr{drive}{i}{}}{\t{}{B_i}} \, + \, \frac{d}{dt} \Vdr{drive}{i}{},  
\end{equation}
is the stimulus driven input to BCs. It takes this form to ensure that, in the absence of ACs coupling, $V_{B_i}(t)=\Vdr{drive}{i}{}(t)$.

CNO binds to specific designed receptors leading either to an excitatory or inhibitory response, depending on the receptors ($G_i$ or $G_q$) \citep{Roth2016, Urban2015}. Here, we propose a simplification of the actual biological process for mathematical convenience. We model the CNO effect by a current of the form $-g_{CNO_T} \, \pare{V - \cE_{CNO_T}}$, where $V$ is the voltage of a cell sensitive to CNO, $T$ is the cell type (i.e. only ACs or RGCs according to the experimental set-up); $g_{CNO_T}$ is the conductance of channels sensitive to CNO that is zero in the absence of CNO while it increases with CNO concentration and $\cE_{CNO_T}$ is the corresponding reversal potential, positive for excitatory CNO and negative for inhibitory.  

 As CNO modifies the membrane conductance it also induces a change of polarization, which is characterized by the parameter $\zeta_A$ for ACs and $\zeta_G$ for RGCs, with a general form (see section \ref{Sec:Generic}):
\begin{equation}\label{eq:zeta}
\zeta_{T}= \frac{\cE_{CNO_T}}{C} \, g_{CNO_T}
\end{equation}
where $T=A,G$ and where $C$ is the membrane capacitance assumed to be the same for all cells. \\

In eq. \eqref{eq:Diff_Syst}, $\tau_{B_i},\tau_{A_j},\tau_{G_k}$ are the characteristic integration times of the BC $i$, AC $j$, RGC $k$. As discussed in the next section and in section \ref{Sec:RestState} (supplementary),
they depend on CNO in a complex, non linear way. 

\ssSec{Parameters reduction}{ParametersReduction}

The model depends on many parameters that constrain the dynamical evolution of the system \eqref{eq:Diff_Syst}. Although some of our results, like the explicit form for RFs \eqref{eq:KG}, are quite general, it is easier, for analytic derivations and for simulations to further simplify these parameters. 

First, we consider that the synaptic weights from ACs to BCs, are controlled by a unique parameter, $w^->0$, $\W{A}{}{B}{}=-w^- \, \Co{A}{}{B}{}$ (inhibitory synapse) where $\Co{A}{}{B}{}$ is an adjacency matrix ($\Co{A}{j}{B}{i}=1$ if the $j$-th AC connects to the $i$-th BC, and $\Co{A}{j}{B}{i}=0$ otherwise). Likewise,
$\W{B}{}{A}{}=w^+ \, \Co{B}{}{A}{}$ with $w^+>0$ (excitatory synapse). We use a simple form of connectivity, where $\Co{B}{}{A}{}=\Co{A}{}{B}{}$ are nearest neighbors adjacency matrices with null boundary conditions. The retina is regularly tiled by different cell types so this approximation is reasonable, although in reality one cell connects to more than 4 neighbors. Our "cells" must actually be considered as "effective" cells with "effective" interactions. In particular, our parameters $w^-, w^+$ correspond to many synaptic contacts. This simplification allows to considerably reduce the number of parameters shaping the synaptic interactions and affords analytic computations (see supplementary).

Following e.g. \citep{berry-brivanlou-etal:99,chen-marre-etal:13,souihel-cessac:21}, the synaptic weight matrices:
\begin{equation}\label{eq:GaussianPooling}
   \W{B}{i}{G}{k}=\Amp{B}{G} \, \, \frac{e^{-\frac{\Dd{B}{i}{G}{k}}{2 \,\sigma_{p}^2}}}{\sqrt{2 \pi} \,\sigma_{p}}, \quad \W{A}{j}{G}{k}=\Amp{B}{G} \, \, \frac{e^{-\frac{\Dd{A}{j}{G}{k}}{2 \,\sigma_{p}^2}}}{\sqrt{2 \pi} \,\sigma_{p}},
\end{equation}
are Gaussian "pooling"  matrices with $\Amp{B}{G}>0$, $\Amp{A}{G}<0$, and
 with $\D{B}{i}{G}{k}$, the $2$-d Euclidean distance between BC $i$ and RGC $k$. The pooling standard deviation, $\sigma_{p}$ was fixed to adapt to the spatial RF of observed cells (see section \ref{Sec:Fit}). \\


In this context, we can further understand the effect of CNO on the model. We assume, for simplicity, that the lattice is so large that we can neglect boundary effects, so that the characteristic times of BCs, ACs, RGCs and their rest state noted $V_B^\ast$ and $V_A^\ast$, $V_G^\ast$, are uniform in space (the general case is discussed in section \ref{Sec:RestState}). We set $\M{B}{}{G}{}=\sum_{i=1}^N \W{B}{i}{G}{k}$ (resp. $\M{A}{}{G}{}=\sum_{j=1}^N \W{A}{j}{G}{k}$). Strictly speaking, these numbers depend on $k$, due to boundary conditions, but we neglect here this dependence.

The characteristic times are then given by:
\begin{equation}\label{eq:CharTimesSimple}
\left\{
\begin{array}{lll}
\tau_B&=&  \frac{\tau_{L}}{1 -\frac{2dw^- \tau_{L}}{\cE_A} (V_A^\ast-\theta_A)}\\
\tau_A &=&\frac{\tau_{L}}{1 + \tau_{L} \pare{ \frac{\zeta_{A}}{\cE_{CNO_A}} +   \frac{2dw^+}{\cE_B} (V_B^\ast - \theta_B)}}\\
\tau_{G} &=& \frac{\tau_{L}}{1 + \tau_{L} \pare{ \frac{\zeta_{G}}{\cE_{CNO_G}} \,+\,   \frac{\M{B}{}{G}{}}{\cE_B} \, \pare{V^\ast_B \, -\, \theta_B} \,+\, \frac{\M{A}{}{G}{}}{\cE_A} \, \pare{V^\ast_A \, -\, \theta_A}}},
\end{array}
\right.
\end{equation}
where $d$ is the lattice dimension, $\tau_L$ the leak characteristic time,  $\cE_B$ and $\cE_A$ are respectively the reversal potential for the synaptic connection from BCs to ACs, and for ACs to BCs (see sections \ref{Sec:Generic} and \ref{Sec:RestState} for detail). The rest states are given by:
\begin{equation}\label{eq:RestStatesSimple}
\left\{
\begin{array}{lll}
V^\ast_B &=&  \frac{4d^2 \tau_{A} \tau_{B} w^{-} w^{+}  \theta_B 
- 2d \tau_{B} w^{-}\tau_A \, \z{A}{} \,+\, 2d \tau_{B} w^{-} \theta_A}{1+4d^2 \tau_{A} \tau_{B} w^{-} w^{+}}\\
V^\ast_A &=& \frac{4d^2 \tau_{A} \tau_{B} w^{-} w^{+}  \theta_A \,-\, 2d \tau_{A} w^{+} \theta_B \, + \, \tau_A \, \z{A}{}}{1+4d^2 \tau_{A} \tau_{B} w^{-} w^{+}}\\
V^\ast_{G} 
&=& \tau_{G} \bra{\M{B}{}{G}{} \, \pare{V^\ast_B \, -\,\theta_B}\,+\,\M{A}{}{G}{} \,\pare{V^\ast_A \, -\, \theta_A}\,+\, \z{G}{}}.
\end{array}
\right.
\end{equation}
Equations \eqref{eq:CharTimesSimple} and \eqref{eq:RestStatesSimple} hold if $V^\ast_B > \theta_B, V^\ast_A > \theta_A$, i.e. the rest state is not rectified. This entails conditions on $\zeta_A$ that we computed explicitly. However, these computations where too lengthy to be included in the paper. The interested reader will find them on the web page  \url{https://team.inria.fr/biovision/mathematicalderivationscno/}. In particular, considering a positive $\zeta_A$ (excitatory CNO), having   $V^\ast_B > \theta_B, V^\ast_A > \theta_A$ require that $\theta_B<0$ and that it is compatible with $\underline{\theta_A=0}$, a condition that we will assume from now on.

On this web page, we also address the question of the variation of $\tau_B, \tau_A, V^\ast_B, V^\ast_A$ as $\zeta_A$ varies. We found that $\frac{\partial V_{B}^\ast}{\partial \zeta_A} =- \frac{2d  \tau_A \tau_B w^{-}}{1+4d^2 \tau_{A} \tau_{B} w^{-} w^{+}} \leq 0$, meaning that BCs get more hyperpolarized when the excitatory effect of CNO on ACs increases, while $\frac{\partial V_{A}^\ast}{\partial \zeta_A} =\frac{\tau_A}{1+4d^2 \tau_{A} \tau_{B} w^{-} w^{+}}  \geq 0$ (ACs get more depolarized when $\zeta_A$ increases). This result is the one qualitatively expected, but note that our modelling is also quantitative.
In strong contrast, the computation of  $\frac{\partial\tau_B}{\partial \zeta_A}$ and  $\frac{\partial\tau_A}{\partial \zeta_A}$ reveals that the sign of these derivative depends on specific domains in the space of parameters $w^-,w^+,\theta_B$, with nonlinear frontiers. That is, even for this reduced model, the behaviour of the characteristic times as CNO increase depends on network parameters in a non trivial way. There are actually domains where these characteristic times \textit{increase} with $\zeta_A$. 

Thus, while a rapid argumentation would predict that, the characteristic times of ACs should increase because the membrane conductance increases as $\zeta_A$ increases, the opposite can happen as well. This is because, as revealed by eq. \eqref{eq:CharTimesSimple}, the characteristic time $\tau_A$ depends on the rest states (eq. \eqref{eq:RestStatesSimple} ) which depend themselves on $\zeta_A$ via the characteristic times $\tau_A,\tau_B$. This results in a two dimensional implicit non linear system studied on the web page \url{https://team.inria.fr/biovision/mathematicalderivationscno/}. \\

As explained in the introduction, our main goal is to confront this modelling to experiments and extract general statements from it. One major difficulty, however, is that we cannot fine tune the parameter $\zeta_A$ (the CNO conductance) in experiments because it is not possible to establish a CNO dose response curve of the changes in retinal activity. Therefore, even under modelling simplifications it is not possible to fit the model from experiments with the non linear equations \eqref{eq:CharTimesSimple} and \eqref{eq:RestStatesSimple}. When fitting experimental results we computed the characteristic integration times of cells in control (CTL) and CNO conditions independently (see section \ref{Sec:Fit}). 
Likewise, we cannot access the rest state of BCs and ACs. 
We fixed the thresholds $\theta_A,\theta_B$ to zero. (Mathematically, $\theta_B$ should be strictly negative, but we can consider a very small absolute value.) Thus, the rest states \eqref{eq:RestStatesSimple} are zero in the absence of CNO. Nevertheless, in experiments, one can observe a small residual polarisation, which changes in the presence of CNO.
That is why we added the polarization parameter, $b_0$, appearing in eq. \eqref{eq:KT}, a huge, but necessary simplification of the entangled equations \eqref{eq:RestStatesSimple}.

Most of the analysis below will be done considering that no rectification takes place, even under light simulation, so that we essentially consider a linear model. This is supported by the paper of \cite{baccus-olveczky-etal:08} showing that the voltage of
bipolar cells is essentially a linear function of the stimulus for white noise. For a more general analysis dealing with rectification please check the supplementary, as well as the discussion section. 

\ssSec{Experimental set-up}{Experimentalsetup}

In our experiments, excitatory DREADDs (hM3Dq) were activated using CNO on RGCs and ACs co-expressing a certain gene (\textit{Scnn1a} or \textit{Grik4}), triggering a calcium release from organelles and thus, leading to increase of intracellular concentration of free calcium. This resulted in membrane depolarisation and higher neuronal excitability. Our experiments suggested that subclasses of ACs and RGCs could be simultaneously sensitive to CNO but we did not observe any evidence of an effect on BCs. 

Detailed experimental details can be found in our recent publication \citep{hilgen-kartsaki-etal:22}. All experimental procedures were approved by the ethics committee at Newcastle University and carried out in accordance with the guidelines of the UK Home Office, under the control of  the  Animals  (Scientific  Procedures)  Act 1986. Recordings were performed on the BioCamX platform with high-density-multielectrode array (HD-MEA) Arena chips (3Brain GmbH, Lanquart, Switzerland), integrating 4096 square microelectrodes in a 2.67 × 2.67 \SI{}{\mm\squared} area and aligned in a square grid with \SI{42}{\um} spacing. Light stimuli were projected onto the retina using a LED projector. Briefly, the projector irradiance was attenuated using neutral density filters to mesopic light levels.
\section{Results}\label{Sec:Results}


In this section we present the theoretical and numerical results based on our retina model. 
We provide only the main conclusions of the mathematical derivations, which are presented in detail in the supplementary section.

\sSec{Model fitting of ganglion cell receptive fields characterized from experimental data}{GCellsRF}

RGC responses emanate from a dynamic balance of synaptic excitation and inhibition, originating from the interactions of BCs and ACs. We believe that such network connectivity gives rise to various response patterns and we show that our model can capture these joint effects, by providing an analytic form of the RF of the cells.
As we demonstrate, this computation provides us an algorithmic way to fit the model parameters to the light responses recorded from mouse RGCs. One can then infer the possible behaviour of ACs and BCs leading to this RGC response, \textit{even if we do not measure them experimentally.}

\ssSec{Mathematical form of the RF of retinal cells}{MathRF}

The results presented below hold for all cell types. Thus, we label cells with a generic index $\alpha= 1 \dots 3N$. BCs have an index $\alpha= 1 \dots N$, ACs have an index $\alpha= N+1 \dots 2N$, RGCs have an index $\alpha= 2N+1 \dots 3N$ and we write $\cX_\alpha$ the voltage of cell $\alpha$. 

The time evolution of the dynamical system in eq. \eqref{eq:Diff_Syst} is controlled by a matrix, $\cL$, called "transport operator" and explicitly written in the supplementary section \ref{Sec:SupMaths}. $\cL$ depends on the connectivity matrices $\W{A}{}{B}{},\W{B}{}{A}{},\W{A}{}{G}{},\W{B}{}{G}{}$ and on all the parameters controlling the dynamics. The form of $\cL$ also depends on the set of rectified cells. In the following, we assume that cells are not rectified i.e. hyperpolarised BCs do not reach the rectification threshold (the rectified case is discussed in the conclusion section and in the supplementary material). Consequently, the dynamical system \eqref{eq:Diff_Syst} is linear.

In this case, the eigenvalues $\lambda_\beta$, $\beta=1 \dots 3N$ and the eigenvectors $\cP_\beta$  of $\cL$ characterize the evolution of cells' voltages. We note $\cP$ the matrix  transforming $\cL$ in diagonal form (the columns of $\cP$ are the eigenvectors  $\cP_\beta$) and $\cP^{-1}$ its inverse. 
  
In this context, we show in the supplementary section \ref{Sec:Derivation_of_X} that the voltage of a cell with index $\alpha$ is the sum of $4$ terms:
\begin{equation}\label{eq:XalphaDriven}
\cX_\alpha(t) = \Vdr{drive}{\alpha}{}(t) \, + \E{drive}{\alpha}{}(t) \, + \,  \E{CNO_A}{\alpha}{} \, + \,  \E{CNO_G}{\alpha}{}, \quad \alpha =1 \dots 3N.
\end{equation}

\paragraph{Stimulus drive.} The first term, $\Vdr{drive}{\alpha}{}(t)$ corresponds to \eqref{eq:Vdrive}, and is non zero for BCs only. It corresponds to the BCs response in the absence of the ACs network. 

\paragraph{CNO effects.} The terms:
\begin{equation}\label{eq:ECNOA}
\E{CNO_A}{\alpha}{}=\zeta_A   \, \sum_{\beta=1}^{3N}\sum_{\gamma=N+1}^{2N}   \frac{\cP_{\alpha\beta}\, \cP^{-1}_{\beta \gamma}}{\lambda_\beta}, \quad \alpha = N+1 \dots 2N;
\end{equation}
and
\begin{equation}\label{eq:ECNOG}
\E{CNO_G}{\alpha}{}= \zeta_G   \, \tau_{G_\alpha}, \quad  \alpha = 2N+1 \dots 3N.
\end{equation}
correspond, respectively, to the impact of CNO on the voltages of ACs and RGCs. 
There are important nonlinear effects hidden in the terms $\frac{\cP_{\alpha\beta}\, \cP^{-1}_{\beta \gamma}}{\lambda_\beta}$ (eq. \eqref{eq:ECNOA}) which depend themselves on the characteristic times \eqref{eq:CharTimesSimple}. 
Thus, the polarization level of ACs and RGCs is not only fixed by the direct effect of CNO on the cell, but is also tuned by entangled network effects. 

\paragraph{Stimulus-network interaction term.} 
In eq. \eqref{eq:XalphaDriven}, the term :
\begin{equation}\label{eq:Edrive}
\E{drive}{\alpha}{}(t)=\sum_{\beta=1}^{3N}\sum_{\gamma=1}^{N} \cP_{\alpha\beta} \, \cP^{-1}_{\beta \gamma}  
\CRep{\beta}{\gamma} \, \int_{-\infty}^t e^{\lambda_\beta(t-s)} \, \Vdr{drive}{\gamma}{}(s)\, ds, \quad \alpha=1 \dots 3N,
\end{equation}
where $\CRep{\beta}{\gamma}=\frac{1}{\t{}{B_\gamma}} + \lambda_\beta$,
corresponds to the indirect effect, via the network connectivity, of the stimulus drive on (i) BCs, for $\alpha=1 \dots N$; (ii) ACs for $\alpha=N+1 \dots 2N$; (iii) RGCs $\alpha=2N+1 \dots 3N$. Thus, this equation describes the response of \textit{all} cells to the stimulus. Especially, it tells us how the direct input \eqref{eq:Vdrive} to BCs  is modulated by the concerted activity of BCs and ACs.

Mathematically, the term \eqref{eq:Edrive} can be interpreted as follows.
The drive (index $\gamma=1 \dots N$) triggers the eigenmodes $\beta=1 \dots 3N$ of $\cL$, with a weight proportional to $\cP^{-1}_{\beta \gamma}$. The mode $\beta$, in turn, acts on the voltage of cell $\alpha=1 \dots 3N$ with a weight proportional to $\cP_{\alpha\beta}$. The time dependence and the effect of the drive are controlled by the integral $\int_{-\infty}^t e^{\lambda_\beta(t-s)} \, \Vdr{drive}{\gamma}{}(s)\, ds$. 
The eigenvalues $\lambda_\beta$ introduce  \textit{characteristic time scales} in the response: exponential decay rate for the real part, frequency of oscillations for the imaginary part, if any. The right eigenvectors (columns of $\cP$) and the left eigenvectors (rows of $\cP^{-1}$) introduce \textit{characteristic space scales} in the response. The easiest case to figure this out is the nearest neighbours connectivity considered in section 5.3.5. Here, eigenvectors are labelled by an index $n$ which actually corresponds to a wave vector, (eq. \eqref{eq:EigenvectorsDelta}), the inverse of a space scale. In this context, the eigenmode with $n=1$ has the largest space scale (the size of the lattice) whereas the eigenmode with $n=N$ corresponds to the minimum space scale (the distance between two nodes). Thus, eq. \eqref{eq:Edrive} illustrates that the response to a spatio-temporal stimulus is a combination of multiple time scales and space scales, leading to interesting phenomena such as resonances as commented below, or waves of anticipation as studied in \cite{souihel-cessac:21}.

\paragraph{The Receptive Field of all cell types.} 
Introducing the function $e_\beta(t) \equiv e^{\lambda_\beta \, t} \, H(t)$ so that $\int_{-\infty}^t e^{\lambda_\beta(t-s)} \, \Vdr{drive}{\gamma}{}(s)\, ds \equiv \bra{e_\beta \conv{t} \K{B}{T} \conv{t} \pare{\K{B}{S_\gamma} \conv{x,y} \cS}}(t)$, and using the separated kernel form \eqref{eq:KT},
the response \eqref{eq:Edrive} reads:
\begin{equation}\label{eq:RFConv}
\E{drive}{\alpha}{}(t)= \bra{\K{}{\alpha} \conv{x,y,t} S}(t),
%
\end{equation}
with:
\begin{equation}\label{eq:KG}
\K{}{\alpha}(x,y,t) = \sum_{\beta=1}^{3N} \pare{\cP_{\alpha\beta} \,  U_\beta(t) \, \times \, \sum_{\gamma=1}^{N} \cP^{-1}_{\beta \gamma} \, \CRep{\beta}{\gamma} \, \K{B}{S_\gamma}(x,y)},
\end{equation}
where we have set $U_\beta(t) \equiv \bra{e_\beta \conv{t} \K{B}{T}}(t)$.
The response of cell $\alpha$ is thus expressed as a convolution of the stimulus with a spatio-temporal kernel $\K{}{\alpha}(x,y,t)$, an expected result from the linear response. Nevertheless, it's important to point out that the expression \eqref{eq:KG} holds for all cell types, not only RGCs and that it contains the network effects  induced by the BCs-ACs network. Thus, 
for $\alpha=1 \dots N$, equation \eqref{eq:KG} characterizes the indirect (network induced) response of BCs to the stimulus drive, in addition to the direct response \eqref{eq:Vdrive}.
For $\alpha=N+1 \dots 2N$, equation \eqref{eq:KG} represents the RF of ACs. Finally, for $\alpha=2N+1 \dots 3N$ we obtain the RF of RGCs. We focus on this last case from now on, essentially because this predicted RF can be confronted to experiments, whereas we have no experimental access to BCs or ACs RF. 

\paragraph{The Receptive Field of RGCs.} 

Henceforth, we will refer to $\K{}{\alpha}(t)$ as $\K{G}{\alpha}(t)$, to make explicit that we are dealing with RGCs. The RF of RGCs can be often written as a product of a space dependent term and a time dependent term (separability).
In our case, this would correspond to write $\K{G}{\alpha}(x,y,t)$ in the form of a product $\K{G}{\alpha}(x,y,t)=\K{G}{T_\alpha}(t) \times \K{G}{S_\alpha}(x,y)$ 
where:
\begin{equation}\label{eq:KGT}
\K{G}{T_\alpha}(t) =   \sum_{\beta=1}^{3N} \cP_{\alpha\beta} \, U_\beta(t),
\end{equation}
 is a temporal kernel and:
\begin{equation}\label{eq:KGS}
\K{G}{S_\alpha(x,y)} =  \sum_{\gamma=1}^{N} \cP^{-1}_{\beta \gamma} \, \CRep{\beta}{\gamma} \, \K{B}{S_\gamma}(x,y),
 \end{equation}
is a spatial kernel.

This separation is not strictly possible in eq. \eqref{eq:KG}, because there is a dependency on $\beta$ on the term $\sum_{\gamma=1}^{N} \cP^{-1}_{\beta \gamma} \, \CRep{\beta}{\gamma}  \, \K{B}{S_\gamma}(x,y)$. Thus, from this computation the RF of RGCs is not expected to be separable in general. However, it is important to remark that the stimulus used in experiments for determining RFs, the white noise, is uniform in probability in the space domain and in the time domain. This induces an effective separation of the response which might not hold for more complex stimuli (e.g. moving objects). As the experimental RFs considered here have been obtained from white noise we will assume separability from now on. 


\paragraph{The spatial part of the RGCs RF.} Equation \eqref{eq:KGS} appears as an overlap of spatial RFs of BCs. 
In such naive overlaps approximations, spatial RFs of BCs are just summed up with a uniform weight. However, here the RF of each BC is constrained by ACs lateral connectivity, via the term $\cP^{-1}_{\beta \gamma}\, \CRep{\beta}{\gamma} $.  In particular, equation \eqref{eq:KGS} is not necessarily circular even if BCs RFs are, and the center of the RGC cell RF is not necessarily at the barycentre of connected BCs RFs. This holds, for example, if AC connectivity is not invariant by rotation. 

\paragraph{The temporal part of the RGCs RF,} \eqref{eq:KGT}. As we consider monophasic temporal kernels $\K{B}{T}$ of BCs with the form \eqref{eq:KT} we have:
%


%
\begin{equation}\label{eq:Ubeta}
\resizebox{.9\hsize}{!}{$U_\beta(t) = A_0 \, \pare{\frac{\bra{ 2 \tau_{RF}^2\, e^{\lambda_\beta \, t} - \pare{t^2 \, \pare{\lambda_\beta\tau_{RF}+1}^2 + 2 t \tau_{RF} \pare{\lambda_\beta\tau_{RF}+1} + 2 \tau_{RF}^2} \, e^{-\frac{t}{\tau_{RF}}} }}{2 \, \pare{\lambda_\beta\tau_{RF}+1}^3 \, \tau_{RF}^2}} \, H(t)$}
\end{equation}
%

$U_\beta$, and, thereby, $\K{G}{T_\alpha}(t)$ is the temporal part of the RGC receptive field, that changes their shape due to variations in the eigenvalues of $\cL$, who are themselves controlled by model parameters. A striking effect arises when some eigenvalues become complex, leading to temporal oscillations of $U_\beta$. This remark is at the core of the analysis exposed in section \ref{Sec:RFSMap}.


\ssSec{Fitting the RFs of ganglion cells}{Fit}


Experimentally, RGCs RFs were reconstructed from Spike Triggered Average (STA) in response to Shifted White Noise (SWN). The SWN, introduced by \cite{Hilgen2017a, Pamplona2021}, is a spatially uniform noise where the images of random black and white squares in the classical "White Noise" stimulus have, in addition, random spatial shifts, improving the resolution of the STA. 
This allowed us to fit the spatial and temporal part of the RF. Note, however, that it is difficult to obtain the surround in the spatial part. From the center part of the spatial RF, we fixed the parameter $\sigma_p$ in the Gaussian pooling, eq. \eqref{eq:GaussianPooling}. 
More extended results can be found in E. Kartsaki thesis \citep{kartsaki:22}, where we display experimental spatial RFs, see e.g. section 3.4.1 of the thesis.

We mainly focused on the temporal part of the RF. The time RF estimation resulted in temporal traces with duration $600$ ms sampled with a rate $33/4=8.25$ ms.
In order to assess the validity of the model, we have fitted these time traces in CTL and CNO conditions. We have 
a data base of $117$ cells sensitive to CNO, i.e. exhibiting increase or decrease in firing rate beyond a certain threshold. 

These reconstructed temporal RFs provide the linear response of a RGC to a spatially uniform flashed stimulus, mathematically corresponding to a Dirac distribution. As we wanted to compare our model's output, the RGC voltage, $V_G$, to this experimental RF, computed from firing rates, we neglected the effect of non linearities and assumed that the experimental response is proportional to the RGC voltage. 
We considered a one dimensional model (chain) with $N=60$ cells of each type where the cells at the boundaries have a fixed, zero, voltage (zero boundary conditions), corresponding to the reference rest state. To reduce the boundaries effect, we made the fit for the RGC in the center of the network.

\begin{figure}
\centerline{
\resizebox{1.2\textwidth}{0.4\textheight}{
\includegraphics[]{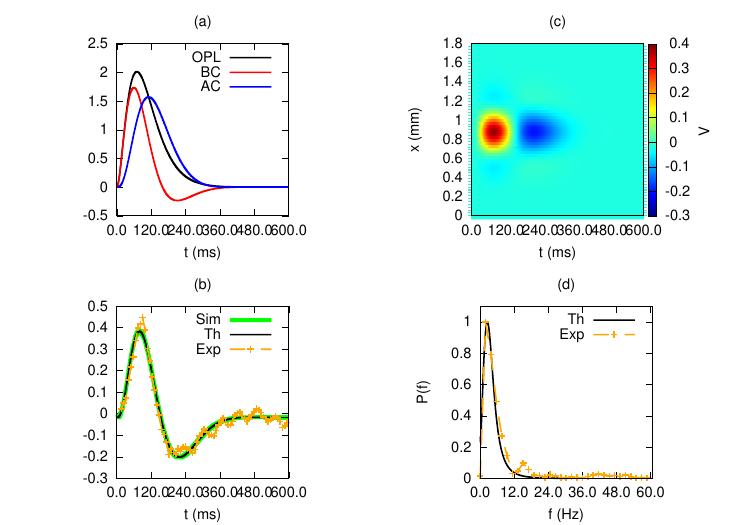} 
}
}
%
%
 \caption{\doublespacing \textbf{Summary panel.} The top left panel \textbf{(a)} illustrates the simulated responses of the OPL (term $\Vdr{drive}{\alpha}{}(t)$ in   \eqref{eq:Vdrive}, black trace), BCs (red trace), ACs (blue trace) connected to the RGC located at the centre of the network. The bottom left panel \textbf{(b)} shows the simulated response of the RGC (green, "Sim", and black, "Th", trace) vs the experimental temporal STA of this cell (orange dots). The green trace ("Sim") is the result of a numerical simulation of the dynamical system \eqref{eq:Diff_Syst} under a spatially uniform flashed stimulus, whereas the black trace ("Th") is the result given by the analytic expression \eqref{eq:KG}.
The top right panel \textbf{(c)} shows the spatio-temporal RF of the RGC, time in abscissa, space in ordinate. The bottom right panel \textbf{(d)} displays the power spectrum of the time response, experimental (orange dots) and theoretical (black lines).
%
 \label{Fig:FitCTL}}
 \end{figure}

We performed simulations of the model \eqref{eq:Diff_Syst} using a spatially uniform Dirac pulse as the stimulus and compute the cell responses, using two modalities: simulation of the differential equations  \eqref{eq:Diff_Syst} (green traces labelled "Sim" in Figure \ref{Fig:FitCTL}) and
analytic computation \eqref{eq:KG} (black traces labelled "Th" in Figure \ref{Fig:FitCTL}).
We observe that these two traces are always identical confirming the goodness of the simulation scheme.

We recall that the parameters shaping this response are:
$A_0$, the intensity of the OPL input;
$b_0$, controlling a residual depolarization/hyper polarization observed in experimental responses;
$w^+$, controlling the synaptic intensity from BCs to ACs; 
$w^-$, controlling the synaptic intensity from ACs to BCs; 
$\Amp{B}{G}$, controlling the synaptic intensity from BCs to RGCs; 
$\Amp{A}{G}$, controlling the synaptic intensity from ACs to RGCs; 
$\tau_B$, the characteristic membrane time scale of BCs;     
$\tau_A$, the characteristic membrane time scale of ACs;     
$\tau_{RF}$, the characteristic membrane time scale of the OPL drive;     
$\tau_G$, the characteristic membrane time scale of RGCs.  

We note $\veta$ the set of all the parameters shaping response. $\veta$ is therefore a point in a $10$-dimensional space.  


The fit was then  done by a gradient descent to minimize the $L^2$-distance $D_2(\veta)$
%
%
between the experimental trace of the time STA, $STA(s)$ and the theoretical temporal RF \eqref{eq:KGT} which depends on $\veta$. 
 The minimization is done by iterating the differential equation:
$$
\frac{d \veta}{du} = - \vnabla_{\veta} D_2.
$$
%
The gradient of $\vnabla_{\veta} D_2$ involves $\vnabla_{\veta} \K{G}{T_\alpha}(s)$ which can be explicitly computed when we have the analytic form of $RF$, or numerically. Note that having the analytic form gives better results especially because it allows second order corrections (Hessian). Our minimisation is done only on the temporal trace. Nevertheless, the simulation allows us to draw the corresponding spatial RF.
\\ 

Although the experimental temporal RFs were quite diverse among cells, we were able to fit all of them with a very good accuracy (final error smaller than $1\%$). We rejected fits where some parameters became unrealistic (e.g. $\tau_A$ larger than $1$ s or $\abs{w^-}>1$ kHz). We rejected about $4\%$ of the fits.  
An example of fit is shown in Figure \ref{Fig:FitCTL}. 

In this synthetic representation the simulated responses of the OPL, BCs and ACs connected to the RGC located at the centre of the network appears in the top left figure. We also show the simulated response of the RGC vs the experimental temporal STA of this cell (bottom left). On the top right we see the numerical spatio-temporal RF using a color map. Finally, at the bottom right, we show the power spectrum (modulus of the Fourier transform) of the temporal response. As developed below, this spectrum provides important information on the cell response.

RGCs RFs in the presence of CNO could be fit equally well with our model.  All fitted cells, in CTL and CNO conditions, can be found on the web page \url{https://gitlab.inria.fr/biovision/dreadds}. The C-code used for simulations can also be found there.
%
%
%

\sSec{Network connectivity shapes the receptive fields of ganglion cells}{Map}

Throughout our analysis of the experimental data, we observed great variability in the effect of DREADD activation with CNO on the RGCs responses. 
In this section, we develop the consequences of our mathematical analysis in an attempt to explain this observed diversity of CNO  effects on RF features.
We propose here an explanation purely based on network effects. There are certainly other possible interpretations based on single cell characteristics such as non linear effects due to changes in conductance etc, discussed in the discussion section. 
The main advantages of our analysis is that it determines network effects on the RGCs RF, controlled by two main parameters, and that it predicts the response to more complex stimuli than full field flashes. 

\ssSec{Two main parameters constrain the RF shape of ganglion cells}{rs}

The entangled, feedback effects of ACs-BCs can be characterized by two dimensionless parameters. 
The first one, $r=\frac{\tau_A}{\tau_B}$, characterizes the ratio between the ACs and BCs membrane integration times. The second, $s=\frac{w^-}{w^+}$, characterizes the ratio between the ACs $\to$ BCs interaction ($w^-$) and the BCs $\to$ ACs interaction ($w^+$).
Of course, the other parameters play an important role when fitting a specific RF. But, what we argue here is that the shape of RF and its space-time scaling essentially depend on the value of $r,s$.

The theoretical explanation is that the RF of a RGC is given by the formula \eqref{eq:RFConv}, which is a cascade of convolutions involving the BC response to the stimulus (OPL input) and the network effects expressed in terms of eigenvalues $\lambda_\beta$ and eigenvectors components $\cP_{\alpha\beta}$ appearing in equation \eqref{eq:KG}. As explained in the supplementary section \ref{Sec:SpectrumL}, these eigenvalues and eigenvectors are essentially tuned by the two parameters $r,s$. There is also a dependence on other parameters discussed in section \ref{Sec:RFSMapExp}. 
Depending on the location in the space $r,s$, some eigenvalues are real, some others are complex.
All eigenvalues have a negative real part, ensuring the stability of the linear system.
Imaginary parts in eigenvalues introduce oscillations in the response, whereas the real part fixes a characteristic decay time. The RF formula \eqref{eq:KG}, involving a sum of exponentials $e^{\lambda_\beta \, t}$ mixes these effects. 
As we considered monophasic OPL response here, the time RF of RGCs is monophasic when all eigenvalues are real. In contrast, oscillations in this RF can appear when some eigenvalues are complex. However, the shape of this RF depends in more detail on the period of oscillations, brought by the imaginary part of complex eigenvalues, and on the characteristic decay times, brought by the real parts.  

\begin{figure}
\centerline{
\resizebox{0.6\textwidth}{0.25\textheight}{
\includegraphics[]{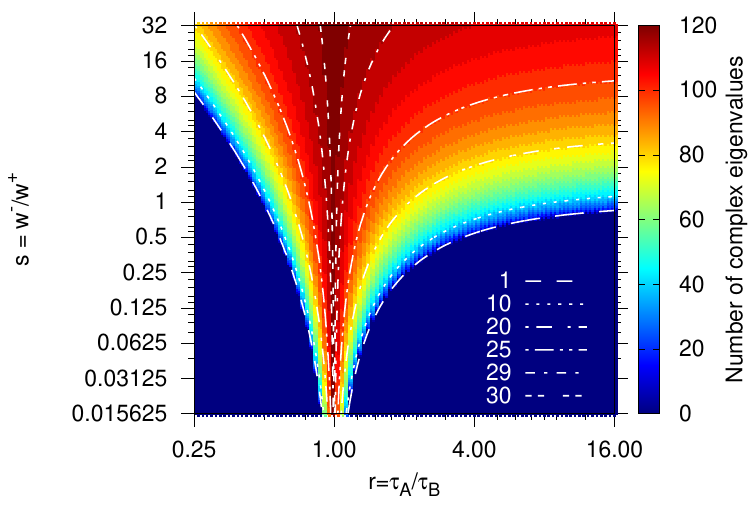} 
}
\hspace{1cm}
\resizebox{0.6\textwidth}{0.25\textheight}{
\includegraphics[]{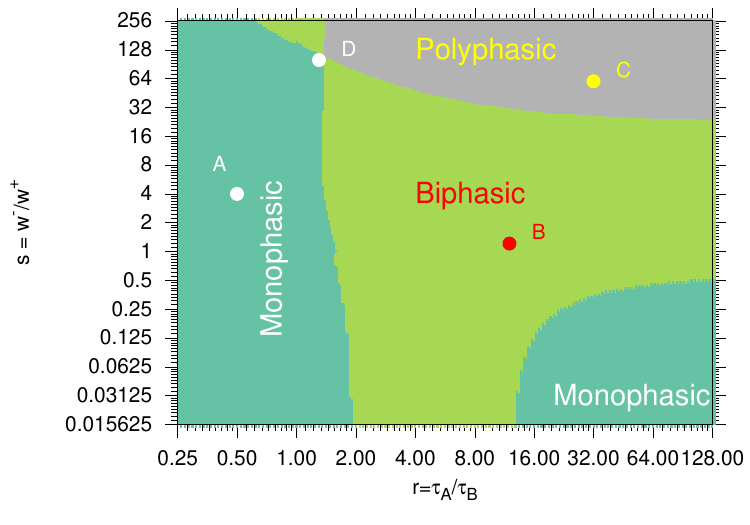} 
}
}
\vspace{0.4cm}
\centerline{
\resizebox{0.5\textwidth}{0.25\textheight}{
\includegraphics[]{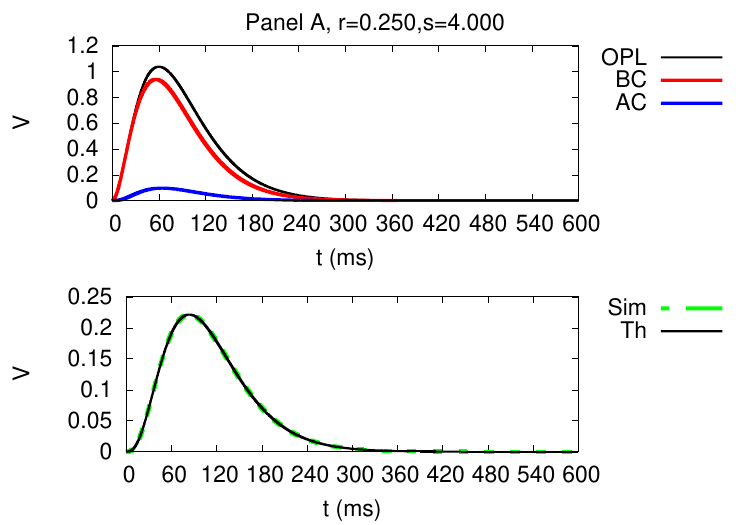}
}
\hspace{0.2cm}
\resizebox{0.5\textwidth}{0.25\textheight}{
\includegraphics[]{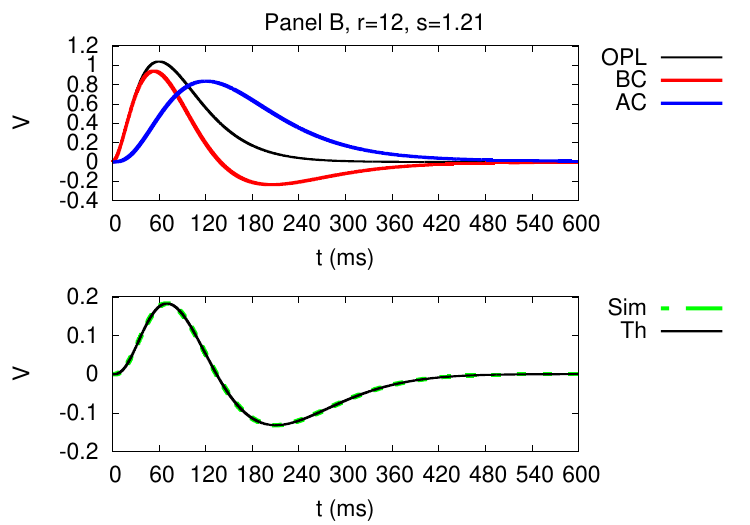}
}
}
\vspace{0.2cm}
\centerline{
\resizebox{0.5\textwidth}{0.25\textheight}{
\includegraphics[]{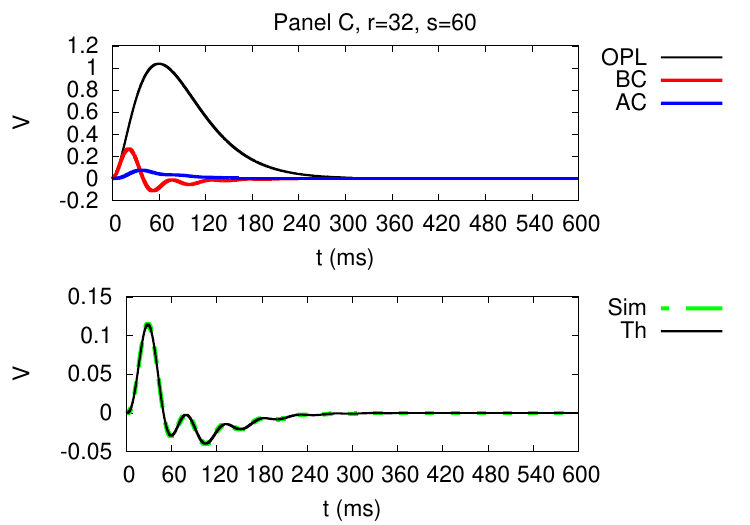}
}
\hspace{0.2cm}
\resizebox{0.5\textwidth}{0.25\textheight}{
\includegraphics[]{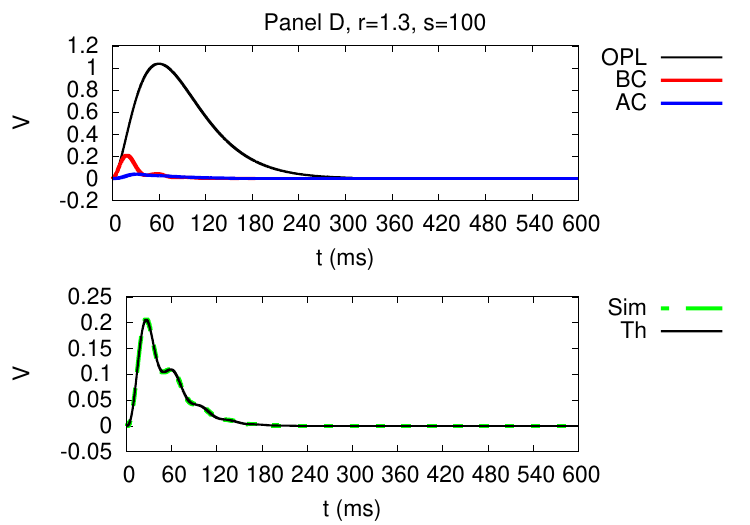}
}
}
 \caption{\small \footnotesize\doublespacing\textbf{Top Left. The skeleton.} Map in the $(r,s)$ plane showing the "skeleton" of eigenvalues structure (white lines). When crossing line $n$ the eigenvalue $n$ switches from real to complex. The color corresponds to the number of complex eigenvalues (see the color bar on the right side).
\textbf{Top right. The RFs Map.} This figures summarizes how the entangled effects of the BCs-ACs network act on the RF of RGCs. In this map in the $(r,s)$ plane we distinguish three main regions (see text for their determination): monophasic, biphasic, polyphasic. The points labelled "A", "B, "C", "D" correspond to the temporal RF plotted in the bottom panels respectively called "Panel A",
 "Panel B", "Panel C", "Panel D". We use the same representation as in Figure \ref{Fig:FitCTL}.
 \label{Fig:PhaseMap}}
 \end{figure}

When moving in the $(r,s)$ plane, the eigenvalue $n$ switches from real to complex conjugate pair when crossing a critical line, depending on $n$, whose equation \eqref{eq:skeleton} is given in the supplementary section \ref{Sec:SpectrumL}. There are $2N$ eigenvalues associated with the BCs-ACs network each one determining a critical line in the plane $(r,s)$. The set of all these lines is what we call the "skeleton". An example of this skeleton is shown in Figure \ref{Fig:PhaseMap}, where we only show some of the critical lines. 
These lines delimit color regions corresponding to the number of complex eigenvalues (see colorbar legend on the right of the figure).

\ssSec{The RFs map}{RFSMap}

The existence of this skeleton determines regions in the $(r,s)$ plane with specific shapes for the temporal RF, given by eq. \eqref{eq:KGT}, a linear combination of functions $U_\beta(t)$ given by \eqref{eq:Ubeta}. The Fourier transform $\hat{U}_\beta(\omega)$ of $U_\beta(t)$ is:
\begin{equation}\label{eq:Ubetahat}
\hat{U}_\beta(\omega)= \frac{1}{\pare{1+i\omega \tau_{RF}}^3}\, \frac{1}{i \omega - \lambda_\beta}.
\end{equation} 
Thus, the Fourier transform of $\eqref{eq:KGT}$ is:
\begin{equation}\label{eq:KGThat}
\K{G}{T_\alpha}(\omega) =   \sum_{\beta=1}^{3N} \cP_{\alpha\beta} \, \hat{U}_\beta(\omega),
\end{equation}
a linear combination of rational fractions. Extending to complex $\omega$s,
$\hat{U}_\beta(\omega)$ has two poles: $\omega= \frac{i}{\tau_{RF}}$ and $\omega=-i \, \lambda_\beta$, corresponding to complex resonances.
%
%
The contributions of all these poles (for $\beta=1 \dots 3N$) are combined in eq. \eqref{eq:KGThat} with weights $\cP_{\alpha\beta}$. 

As we move in the $(r,s)$ plane, we notice the following. When $(r,s)$ are small, eigenvalues are real and the terms $\cP_{\alpha\beta}$ are close to diagonal. In this case, the dominant pole contribution in \eqref{eq:KGThat} is the pole $\omega=i \frac{1}{\tau_{RF}}$ corresponding to the OPL contribution. Equation \eqref{eq:KGThat} has a single peak centered at $\omega=0$, corresponding to a monophasic response. For larger values of $r,s$ some eigenvalues become complex, giving potential additional peaks in the power spectrum. Actually, we observed two cases mutually compatible. First, the central peak at $\omega=0$ switches to a non zero value. This corresponds to the appearance of an exponentially damped oscillation in the RF, giving a biphasic response. However, secondary peaks may appear leading to residual oscillations, in addition to the main trend (monophasic or biphasic). This gives what we call a \textit{polyphasic} response. Such residual oscillations were observed in our experiments and were relatively numerous (about $40 \%$). There are, of course, other hypotheses explaining these residual oscillations, but here, we will support the hypothesis that they are generated by a network effect. An example is given in Figure \ref{Fig:FitCTL} where we observe, at the bottom left, residual oscillations after the main biphasic response, and, at the bottom right, the power spectrum with a main peak not centered at zero and a secondary peak corresponding to the residual oscillations. Note that this secondary peak is observed on experimental data but we failed to reproduce it in the fits. This is further explained in the discussion section. \\
%

This analysis leads us to broadly decompose the $(r,s)$ plane into $3$ regions corresponding to cells response phases: monophasic, biphasic, polyphasic. One switches from one phase to the other when some peaks in the power spectrum appear or disappear, driven by the spectrum of $\cL$. 

\begin{figure}
\centerline{
\resizebox{0.5\textwidth}{0.25\textheight}{
\includegraphics[]{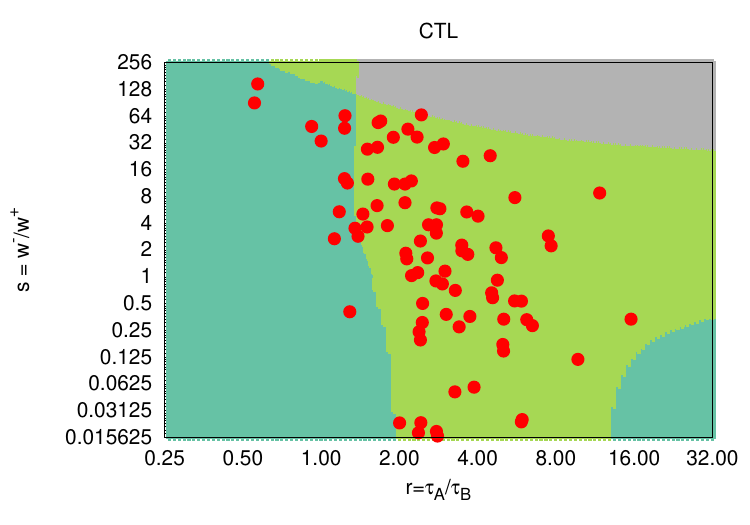} 
}
\hspace{1cm}
\resizebox{0.5\textwidth}{0.25\textheight}{
\includegraphics[]{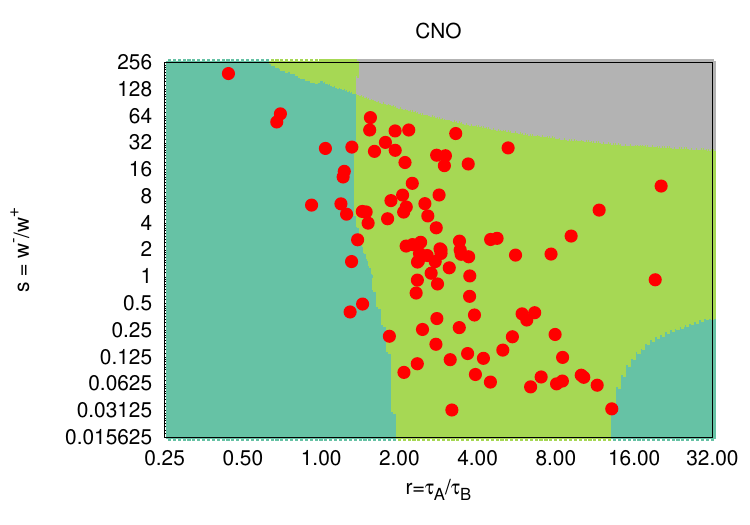} 
}
}
\vspace{0.5cm}
\centerline{
\resizebox{0.4\textwidth}{0.2\textheight}{
\includegraphics[]{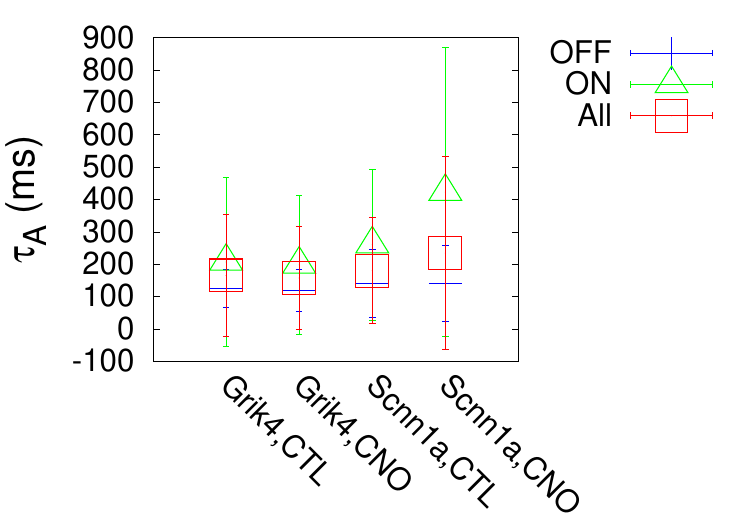} 
}
\hspace{0.5cm}
\resizebox{0.33\textwidth}{0.2\textheight}{
\includegraphics[]{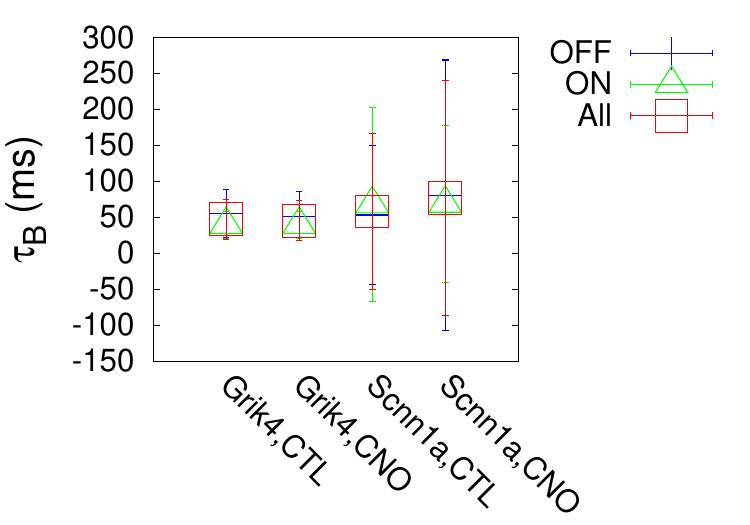} 
}
\hspace{0.5cm}
\resizebox{0.33\textwidth}{0.2\textheight}{
\includegraphics[]{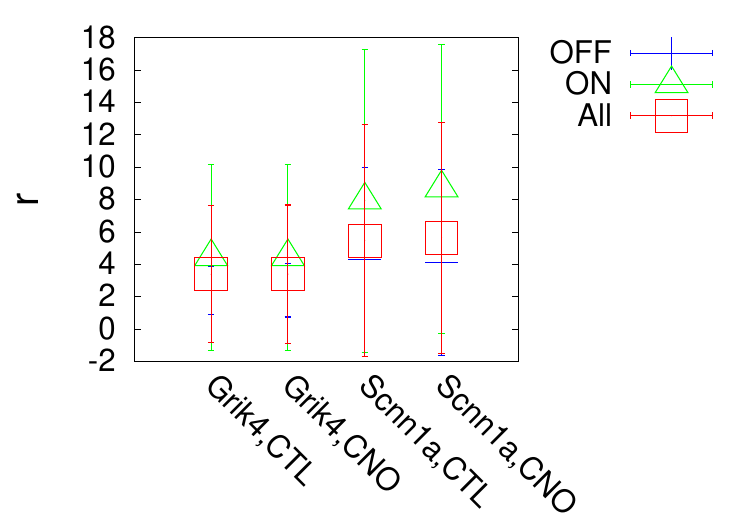} 
}
}
\vspace{1cm}
\centerline{
\resizebox{0.33\textwidth}{0.2\textheight}{
\includegraphics[]{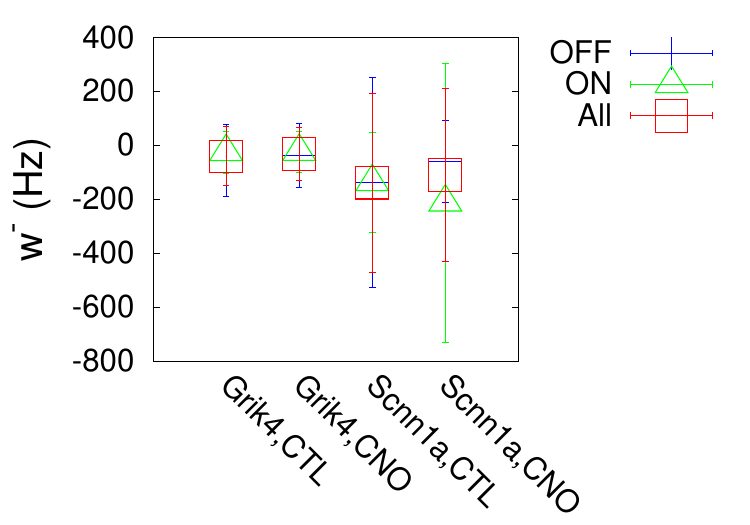} 
}
\hspace{0.5cm}
\resizebox{0.33\textwidth}{0.2\textheight}{
\includegraphics[]{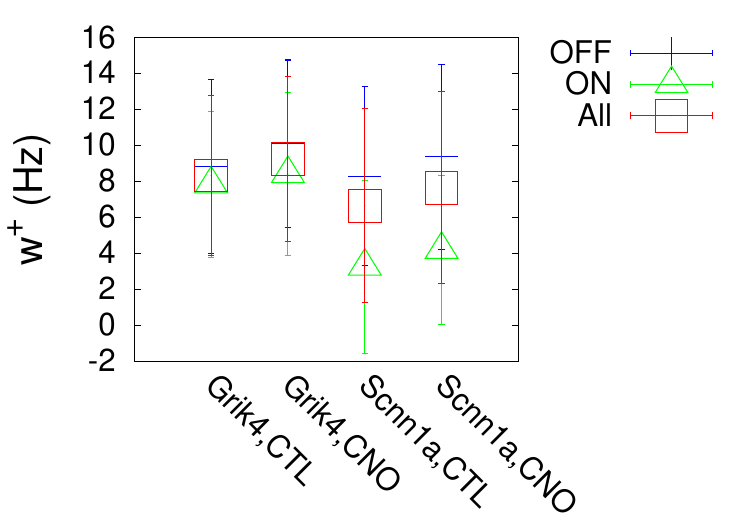} 
}
\hspace{0.5cm}
\resizebox{0.33\textwidth}{0.2\textheight}{
\includegraphics[]{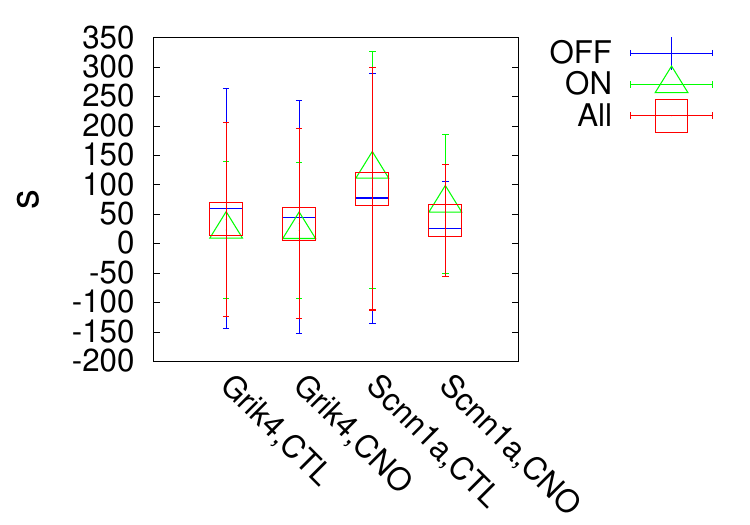} 
}
}
 \caption{\small\doublespacing\textbf{Top. Repartition of fitted experimental cells in the RFs map.}
 Each point corresponds to an experimental cell.
 \textbf{Top left.} CTL conditions. \textbf{Top right.} CNO conditions 
 \textbf{Middle.} Mean and standard deviation of $\tau_A$ (left), $\tau_B$ (center), $r$ (right), fitted from experiments, for genes Grik4 and Scnn1a, in CTL and CNO conditions. We have separated the estimation for OFF cells (blue), ON Cells (green) and all cells (red). 
  \textbf{Bottom.} Mean and standard deviation of $w^-$ (left), $w^+$ (center), $s$ (right), fitted from experiments. The representation is the same as the previous row.
 \label{Fig:PhaseMapExp}}
 \end{figure}

The corresponding "phase diagram", obtained numerically, constitutes what we call the "RFs map" shown in Figure \ref{Fig:PhaseMap} (top right). There are four points, labelled A,B,C,D, on this map, each representing a different cell's response phase. For each point, we have plotted the RGC temporal RF, as computed with the model (bottom panels).
A more general representation of what is going on when moving along a specific pathway in this map can be found at the web page \url{https://team.inria.fr/biovision/cno_paper_supplementary/}, where one can see movies showing how the network effects shape the RGCs RFs when $(r,s)$ vary.  It is important to note that this map is a projection from a $10$ dimensional space in $2$ dimensions. It has been obtained, from the skeleton, by fixing the other parameters values, based on the average values obtained from fits over the cell populations (see e.g. Fig. \ref{Fig:PhaseMapExp}). 

 Figure \ref{Fig:PhaseMap} illustrates how the BCs-ACs network shapes the RF of a RGC by a subtle balance between BCs-ACs, BCs-ACs interactions (parameter s) and the time scale of their response (parameter r). For simplicity, we consider here ON BCs, but the explanation holds also for OFF BCs.

The OPL drive (black line) induces a depolarisation of BCs (red line) within a time scale of order $\tau_B$. This excites the connected ACs (blue line), with an intensity $w^+$. The excitation of ACs hyperpolarises BCs with an intensity $w^-$ within a time scale of order $\tau_A$.  RGCs receive a combination of excitatory and inhibitory inputs from their afferent circuit with respective weights $\Amp{B}{G}$ and $\Amp{A}{G}$.

In the monophasic region, ACs respond in the same time scale as BCs. One observes, for RGCs a monophasic response (Panel "A", middle left) whose intensity depends on the ratio between excitation, provided by BCs, with a weight $\Amp{B}{G}$ and inhibition, with a weight $\Amp{A}{G}$. When moving to the biphasic region, ACs respond slower than BCS, leading to the biphasic response illustrated with panel "B". When moving upward in the RFs map (increasing $s$) this biphasic response becomes polyphasic with oscillations. This is illustrated in panel "C". BCs start to raise due to the OPL drive, leading to a rising of ACs, slower than BCs, leading to a hyper-polarisation of BCs. This leads to a decrease of ACs voltage, thereby, to a rising of BCs which still respond to the OPL drive. This cycle can be repeated several times, depending again on the parameters $r,s$. Note that the amplitude is always decreasing exponentially fast. The period of the observed oscillations and the damping characteristic time depend on the location in the Map. Finally, panel "D" is the point at the intersection of the $3$ phases regions. We show it for completeness. 
Note that increasing $s$ leads to a decrease of the RGCs response. When $s$ is too high, the response becomes too weak to be observed experimentally. 

\ssSec{Experimental cells spread in the RFs map}{RFSMapExp}

To confront our theoretical insight with experiments, we have placed the recorded cells in the RFs map as shown in Figure \ref{Fig:PhaseMapExp}. That is, for each experimentally recorded RGC, we fit the model parameters $\veta$ as explained in section \ref{Sec:Fit} thereby providing an estimation of $r,s$, independently in CTL and in CNO conditions. 
This defines a virtual network, made of identical cells in each layer, where all RGCs are responding like the experimental RGC. As mentioned above, the map is \textit{only a projection} of $\veta$, which exists in a $10$ dimensional space, in the two dimensional plane $r,s$. Some parameters are linked together though. The mathematical analysis in the supplementary section \ref{Sec:SpectrumL} shows us that the skeletons obtained for a fixed value of $\tau_B,w^+$, can be extrapolated to other values ${\tau'}_B,{w'}^+$ by the simple rescaling  $r'=r,s'=\pare{\frac{{w'}^+ \, {\tau'}_B}{w^+ \, \tau_B} }^2 \, s$. 
Using this rescaling, the map of Figure \ref{Fig:PhaseMapExp} has been drawn for a specific value of $w^+=8.5$ Hz and $\tau_B=30$ ms.  This corresponds to mean values of these parameters, averaged over the set of experimental cells (in CTL conditions). In this two dimensional representation, extra information coming from the other parameters $\tau_G,\Amp{B}{G},\Amp{A}{G},A_0,b_0$ is lost.


Figure \ref{Fig:PhaseMapExp}, left, shows us the repartition of cells in the map, in CTL conditions. A few of them are monophasic, but many of them are biphasic, with a significant proportion close to the polyphasic region and showing residual oscillations. The figure \ref{Fig:PhaseMapExp}, right, shows the same cells in CNO conditions. 
The main observation is that CNO (right panel), does not dramatically change the repartition of cells in the RFs map. This is made more explicit in the bottom panels of figure \ref{Fig:PhaseMapExp}. We show the mean and standard deviation of the main model parameters: $\tau_A,\tau_B, w^-,w^+,b_0$ in CTL and CNO conditions, separating the two subclasses of investigated genes: Grik4 and Scnn1a, and separating ON or OFF cells. These parameters are essentially constant showing that there is no statistical trend induced by CNO. 
%
This is in agreement with our previous paper \citep{hilgen-kartsaki-etal:22} where we showed that Scnn1a or Grik4 groups actually include
multiple anatomical cell types, which suggest that our model or fitting does not introduce any obvious bias.

%

\begin{figure}
\centerline{
\resizebox{0.5\textwidth}{0.3\textheight}{
\includegraphics[]{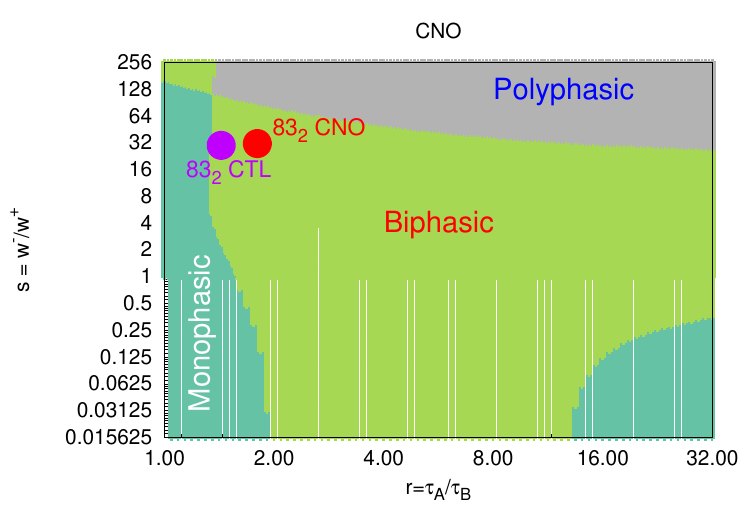} 
}
\hspace{1cm}
\resizebox{0.6\textwidth}{0.3\textheight}{
\includegraphics[]{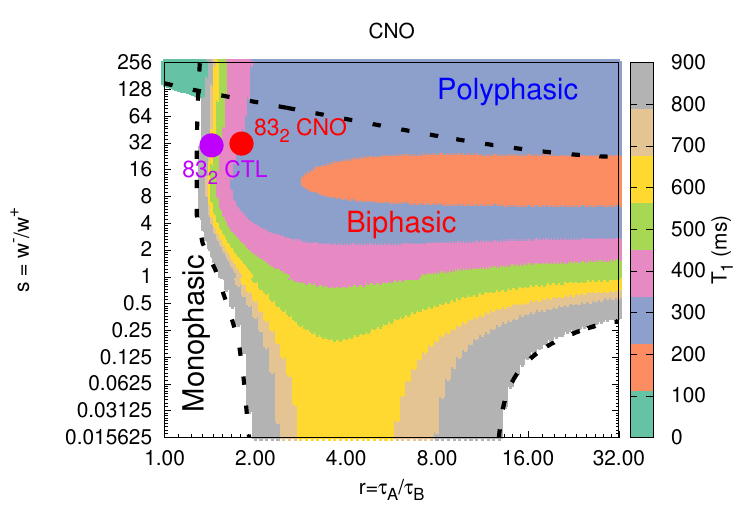} 
}
}
\vspace{1cm}
\centerline{
\resizebox{0.5\textwidth}{0.35\textheight}{
\includegraphics[]{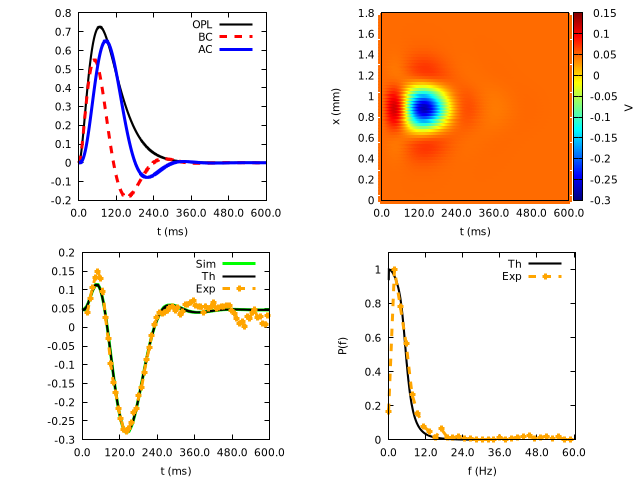} 
}
\hspace{1cm}
\resizebox{0.5\textwidth}{0.35\textheight}{
\includegraphics[]{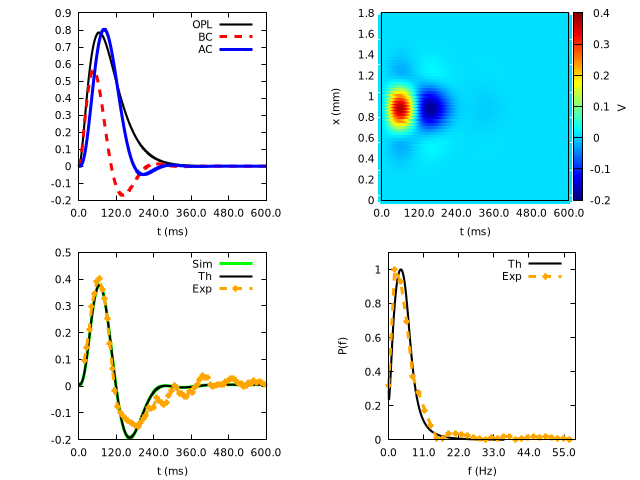} 
}
}
 \caption{\footnotesize\doublespacing\textbf{CNO may change the cell response.} The \textbf{upper left} panels shows the representative point of a cell, labelled $83_2$, in the RF map, in CTL (purple point) and CNO (red point) conditions. This shows the displacement of the representative point of the cell when CNO is applied. The \textbf{upper right panel} shows the same motion, in the same space $(r,s)$, but considering here the main peak in the power
spectrum, corresponding to a characteristic period $T_1$. The dashed black lines correspond to the separation between the monophasic, biphasic, and polyphasic regions. } One switches from a region with a high $T_1$ of order $600$ ms (green area) in CTL, to a region where $T_1$ is of order $300$ ms (grey area) in CNO conditions. As shown in the \textbf{bottom panel} this displacement corresponds to a switch from monophasic to biphasic region.
 \label{Fig:Cell_83_2}
 \end{figure}
\begin{figure}
\centerline{
\resizebox{0.5\textwidth}{0.3\textheight}{
\includegraphics[]{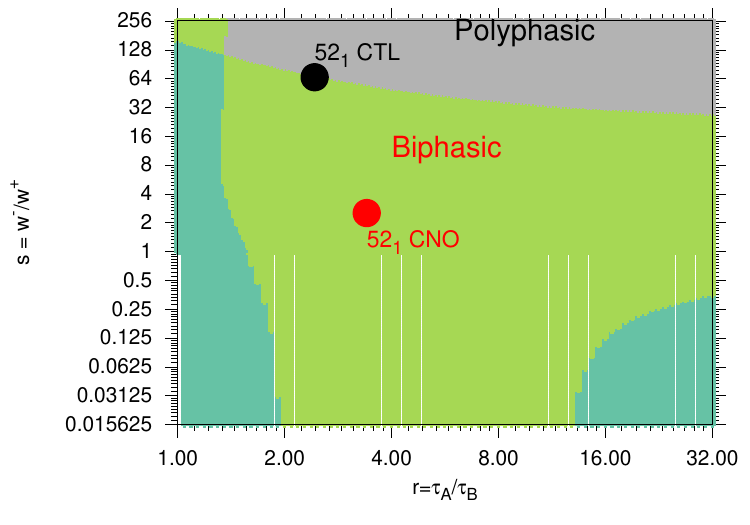} 
}
\hspace{1cm}
\resizebox{0.6\textwidth}{0.3\textheight}{
\includegraphics[]{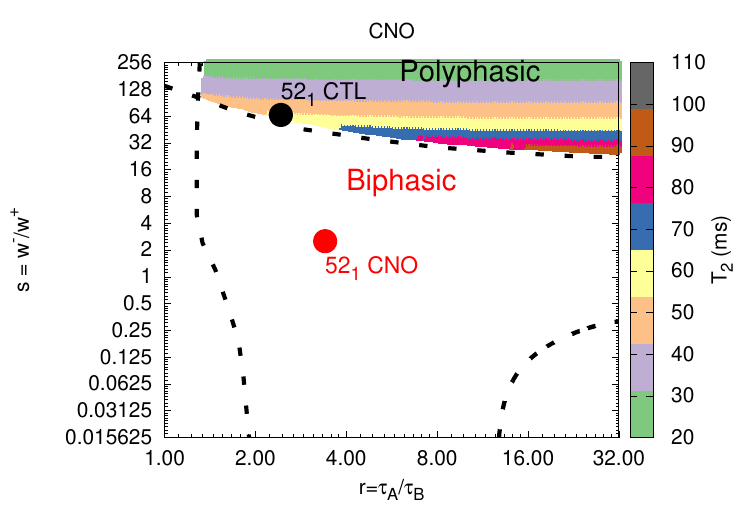} 
}
}
\vspace{1cm}
\centerline{
\resizebox{0.5\textwidth}{0.35\textheight}{
\includegraphics[]{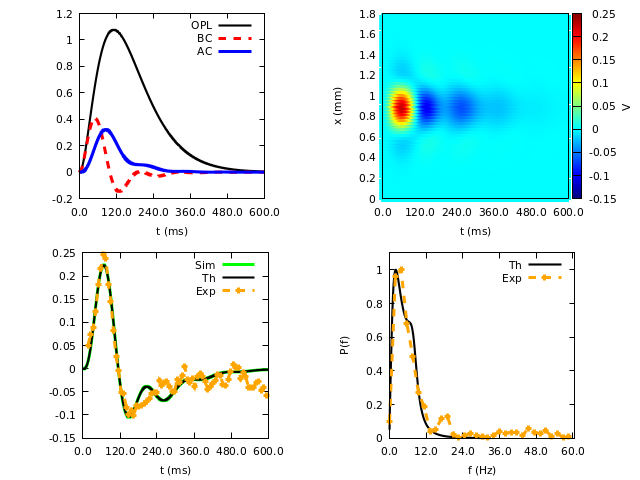} 
}
\hspace{1cm}
\resizebox{0.5\textwidth}{0.35\textheight}{
\includegraphics[]{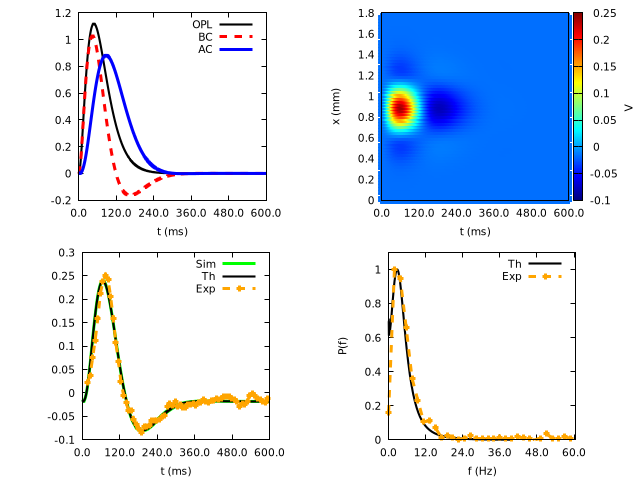} 
}
}
 \caption{
\doublespacing \textbf{CNO may change the cell response.} \textbf{Left top.} Application of CNO moves the representative point of the cell (here label $52_1$) in from  polyphasic to biphasic region resulting in a change in the cell response (\textbf{Bottom panel}). The figure \textbf{Right top} shows how the period of polyphasic oscillations depends on the place in the RFs map. 
 \label{Fig:Cell_52_1}}
 \end{figure}

The situation is radically different when investigating the effect on \textit{individual} cells. Indeed, the application of CNO makes some cells to move their representative point from one region in the RF map to the other, thereby drastically changing the cell's response. Two examples are shown in Figure \ref{Fig:Cell_83_2} and \ref{Fig:Cell_52_1}. In Figure \ref{Fig:Cell_83_2}, the application of CNO induces a motion of the representative point in the RFs map.  However, as the cell is close to the area separating the monophasic from the biphasic phase, this motion impacts dramatically the shape of the time response. 
Note that this motion looks tiny because it is plotted in log scale, but in reality reflects a relatively large change in the RGC properties. This corresponds therefore to a relatively big change in the properties of the ganglion cell. Actually, the RFs map can be refined by plotting the value of the main period $T_1$ (corresponding to the main peak in the power spectrum) as shown in the top right figure. The dashed black lines correspond to the separation between the monophasic, biphasic, and polyphasic regions. Note that the switch from bi- to poly-phasic corresponds to additional peaks in the spectrum that are not visible when considering the period of the first peak, $T_1$. (This transition is seen in  Fig. \ref{Fig:Cell_52_1} where the period of the second peak is shown). In CTL conditions, the cell is located in  the green region with a high $T_1$ of order $600$ ms, in the limit of experimental resolution. 
With CNO, the cell switches to the grey region where $T_1$ is of order $300$ ms. At the bottom of the figures, the synthesis panel (same representation as in Figure \ref{Fig:FitCTL}) for CTL (left) and CNO conditions (right) is presented.


%

In Figure \ref{Fig:Cell_52_1} we show a motion inducing a switch from polyphasic to biphasic. 
Here, CNO drives the cell from the boundary of the polyphasic region to the biphasic region. The representation is the same as for Figure \ref{Fig:Cell_83_2} except that the top right figure displays the period of the second period $T_2$ (secondary peak in the power spectrum). 

 \medskip
 
 \textbf{Remark.} The range of parameters {r, s} is very broad, and seems to be stretched in terms of biological plausibility. In Fig. 5, $s$ can raise up to $s=200$  which means that ACs inhibitory feedback is $200$ times stronger than BCs excitation onto ACs which looks quite implausible. In the same vein, we observe values of $r = 30$ corresponding to ACs having a $30$-fold longer time constant than BCs, which might be questionable.  These values must actually  not be taken literally as they correspond to several artifacts. First, as explained above, $s$ has been rescaled to match a RF map drawn from a set of reference parameters with e.g. $w^+=8.5$ Hz, and $\tau_B=30$ ms. In addition, as mentioned earlier, these synaptic weights are actually effective interactions that mimic the effect of cells populations.  These extreme values of $(r, s)$ are therefore somewhat reflecting the fact that we are dealing with interactions among multiple cell types in the inner retina.

\section{Discussion}\label{Sec:Conclusions}

In this paper, we investigated the role of AC-mediated lateral connectivity in the response of RGCs to visual stimuli. 
Our conjecture was that these responses are strongly constrained by such lateral connectivity. Based on the mathematical analysis of a network featuring the interaction of BCs-ACs-RGCs we were able to produce an analytic form for the spatio-temporal response (receptive fields) of all cell types in the model. This finding has significant implications for the usefulness, identifiability (i.e. its parameters could be obtained from experimental data) and interpretability of the model. First, it provides an algorithmic way to fit the model parameters to the light responses recorded from mouse RGCs, using the analytical formula of the RF. This means that we are able not only to find the parameters that best fit the variables concerning the RGCs responses, but also to infer the possible behaviour of ACs and BCs leading to the RGCs responses, even if we don’t measure them experimentally. Second, it provides an intuitive understanding of the role of various model variables and highlights the impact of two phenomenological parameters (with a physical meaning) on the spatio-temporal response (i), the intensity of the interactions BCs-ACs, and, (ii), the characteristic time scale of these cells responses. This can be summarized in the two dimensional RF maps, where one observes phases corresponding to different modalities in the response. We were able to validate experimentally these modelling results, based on the ability to pharmacologically modify the level of ACs and RGCs neural activity using pharmacogenetics (DREADD-CNO). We would like now to comment some caveats and potential extensions of this work.

\paragraph{Polyphasic phase.} Although we observe about $40 \%$ of polyphasic cells in the experimental plots (characterized by secondary peaks in the power spectrum) the model has difficulty to properly fit them. This is visible in figure \ref{Fig:PhaseMapExp} where no cell is within the polyphasic region whereas the secondary peaks are clearly visible in the experimental power spectra (see Figure \ref{Fig:Cell_52_1}). This can be explained by several factors. First, the fitting method, trying to estimate a model with $10$ parameters from a trace with a few hundreds of points, has clearly limits. In particular, the secondary peaks, having a few points in the power spectrum, are hard to capture and require a patient fine tuning. The main limitations may also come from the model itself, as developed in the following. 


\paragraph{Inhomogeneities.}  Our model is quite homogeneous. It assumes that there is only one subclass for each cell type, represented by a unique set of parameters, the connectivity is fairly regular ... This homogeneity was the key to answer a specific question:  understanding better  the potential role of the BCs-ACs network on the RGCs response, both from a theoretical and experimental perspective. This study was actually done in the spirit of theoretical physics: find a set of canonical equations, grounded on reality (here biophysics), and mathematically analyse their behaviour with a \textit{minimal} set of reduced (dimensionless) parameters. Then, confront the model predictions to experiments and propose a simple representation of the observed effects. This type of approach is useful, because it allows one to understand the action of cells within a large retinal network at a level of detail and comprehensiveness generally not possible by numerical simulation alone. Yet, one may wonder what would happen if  more parameters were taken into consideration, to get closer to real retinal networks. For example, what would be the consequences of  introducing more sustained and transient bipolar and amacrine cells ? Or considering more general forms of connectivity, e.g.  connections between amacrine cells ?

As mentioned at the end of section \ref{Sec:Structure}, equation \eqref{eq:Diff_Syst} extends to more cell types and general connectivity. The same holds for equations \eqref{eq:Edrive} and \eqref{eq:KG}. For example, considering 
$M_B$ types of BCs, each with a specific RF $\cK_m(x,y,t)=\K{B}{S_m}(x,y)\, \K{B}{T_m}(t)$,  equation \eqref{eq:KG} would extend to: 
\begin{equation} \label{eq:KG_mult_pop}
   \K{}{\alpha}(x,y,t) = \sum_{\beta=1}^{3N} \pare{\cP_{\alpha\beta} \,  \sum_{m=1}^ {M_B} U_{\beta m}(t) \, \times \, \sum_{\gamma \in m} \cP^{-1}_{\beta \gamma} \, \CRep{\beta}{\gamma} \, \K{B}{S_\gamma}(x,y)},
\end{equation}
where $U_{\beta m}(t) \equiv \bra{e_\beta \conv{t} \K{B}{T_m}}(t)$ and where $\gamma \in m$ stands for "the BC $\gamma$ belongs to population m". 
Beyond that, we are, at the moment not able to tell what would be the impact of such heterogeneity on the model. The reason is fairly simple. Adding heterogeneity will increase the number of parameters, 
and, in general, will prevent us to go as far as we did in this paper, constructed in the spirit of having a mathematical control of the results. On technical grounds,  the structure of eigenvalues and eigenvectors is expected to dramatically change in a way that we cannot, in general, determine analytically, in contrast to the present work, and it would require numerical explorations in a huge parameters space.

Adding heterogeneity clearly necessitates to precisely define which question we want to answer, and, from this point of view, \textit{specific} forms of heterogeneity could be studied in an extension of this work. A simple idea would be to introduce some randomness in the parameters characterizing BCs and ACs. For example, considering random leak times in order to have a distribution of characteristic times $\tau_A, \tau_B$ to study e.g. how we depart from the result obtained here when increasing the variance of these times. Or, considering random versions of the incidence matrices $\Co{A}{}{B}{}, \Co{B}{}{A}{}$. An example of this has been considered in the paper \cite{souihel-cessac:21} to analyse anticipation effects. In any case, such questions would require long investigations beyond the scope of this paper.

\paragraph{Rectification.} The model includes weak non linearities (rectification) that were neglected in the mathematical computations. The effects of such rectification can be mathematically investigated \citep{cessac:22}. Mainly, rectification projects dynamics on the subspace of non rectified cells. This means that the dimensionality of the dynamical system changes in time, depending on the stimulus and network parameters, with strong consequences on the spectrum of $\cL$, and thereby, on the power spectrum briefly discussed in the supplementary section \ref{Sec:SpectrumL}. 

\paragraph{Local non linearities vs network effects.} Additional non linearities take place in retinal dynamics. Ion channels have a non linear behaviour inducing phenomena such as bifurcation and bursting, essential, for example, in the development of the retina were bursting Starburst Amacrine Cells generate  retinal waves \citep{hennig-adams-etal:09,karvouniari-gil-etal:19,cessac-matzakou-karvouniari:22}. In addition, gain control plays also a central role in the response to spatio-temporal stimuli inducing, for example, retinal anticipation \citep{berry-brivanlou-etal:99,chen-marre-etal:13,souihel-cessac:21}. 


Although our study was limited to responses to full-field flashes we would like to extend the consequence of our analysis to more complex stimuli. First, the presence of peaks in the power spectrum implies the existence of \textit{resonances}, that is preferred frequencies for the RGCs. Exciting a cell with a resonance frequency will produce a maximal response. When applying a stimulus like the Chirp stimulus \citep{hilgen-kartsaki-etal:22, Baden2016} there is a phase where periodic flashes, with constant contrast but increasing frequency, are applied.
One observes frequently a bump in the experimental RGCs response that might correspond to such a resonance. 

In addition to preferred time frequencies, our analysis also shows that the response of RGCs, induced by the network, may also involve specific \textit{space} scales. Mathematically, these scales appear in the eigenvectors of the transport operator (see eq. \eqref{eq:EigenvectorsDelta} in the Supplementary section \ref{Sec:Nearest_neighbours}). The practical implication would be that, presenting a localised, time periodic, stimulus at a resonant frequency and with small radius,  and increasing slowly this radius, one may observe scales where the response is maximal. One of these scales may correspond to the size of the RF but we conjecture that there should be other, larger, scales where this phenomenon appears. This would actually be a way to disentangle intrinsic responses of cells, from network induced responses, by blocking the ACs synapses (e.g. strychnine for glycinergic cells). More generally, the existence of time resonances and preferred space scales would also induce resonant response to moving objects with the appropriate speed. Such effects could also be related to mechanisms giving rise to anticipatory waves \citep{menz-lee-etal:20}. \\


To conclude, this research 
provides mathematical insights to explore the potential role of the amacrine cells network in vision processing. First, it brings in the field of retinal modelling methods and concepts from dynamical systems theory, especially, the idea that concerted population activities can be 
understood with a few reduced parameters, still reproducing experimental observations. Second, it makes a step forward to formalizing the concept that the response of RGCs to stimuli is not only the result of intrinsic cells characteristic (e.g. their morphology), but it also depends of their interaction with other cells, and on the stimulus itself. 

This paper was a first step to confront this approach to experiments, as a proof of concept opening up several research tracks.  First, it could be used to infer the potential structure of spike correlations generated by a RGCs population when responding to a moving object. While it is easy to show, in this model, that lateral inhibition decorrelates the response to full field stimuli, the situation is dramatically different when an object is moving in its visual field. The resulting wave of activity 
generates non trivial, transient, spatio-temporal correlations which may contain fundamental information deciphered by the cortex. The corresponding non stationary spike train distributions can be mathematically \citep{cessac:11,cessac-rostro-gonzalez-etal:09,cessac-ampuero-etal:21} and numerically \citep{cessac-kornprobst-etal:17} studied.  
Pushing forward this formalism could therefore be a key toward understanding better how the retina encodes the visual information in a visual world made of motion. 

Second, with the necessity of improving their biological plausibility (see discussion above), this class of model could be used to fit online experimental retinal responses to a simulator, by adapting the stimulus to the model prediction, in a closed loop process \citep{benda-gollisch-etal:07}. This is in the spirit of current trends of research attempting to construct data driven models of the retina \citep{schroder-klindt-etal:20}, although our approach targets to reduce as much as possible the number of parameters and maximize the mathematical control so as to avoid the "black box" effect.


Finally, our work could be used in future studies to explore the role of other RGC subclasses or other retinal neurons and their interactions. In addition, it could be used to disassemble the components of other retinal circuits, by manipulating the activity of specific neurons. It could also potentially benefit research in other parts of the nervous system, as fundamental properties of the inner retina are shared with other parts of the brain.
\section*{Acknowledgments}
This project was funded by the Leverhulme Trust (RPG-2016-315 to ES and BC), by
Newcastle University (Faculty of Medical Sciences Graduate School).


\section*{Supplementary}\label{Sec:Supplementary}


\Sec{Derivation of the model equations and properties}{SupDerivation}

\sSec{Generic equations for voltages}{Generic}

We start from the fundamental equation of neuronal dynamics (charge conservation) and make several simplifying assumptions.
 
First, we assume that neurons are reduced to points. They all have the same leak conductance, $g_L$, the same leak reversal potential, $\cE_L$ and the same membrane capacity, $C$. They are passive (no active ion channels). 
This last assumption is based on the fact that most of the retinal neurons considered here (BCs and ACs) are not spiking. From these assumptions we get rid of equations for activation-inactivation  variables present in most spiking models \citep{ermentrout-terman:10}. For RGCs, spiking is ruled by a LN type model (see main text).

As exposed in the main text we model the CNO effect by a current of the form $-g_{CNO_T} \, \pare{V_q - \cE_{CNO_T}}$ where $T$ is the cell type (e.g. ACs or RGCs) so that the CNO conductance is constant and depends only on the cell type.  $\cE_{CNO_T}$ is the corresponding reversal potential. We insist, however, that this is only a modelling short-cut allowing to mathematically treat inhibitory and excitatory effects on an equal footing, with only one parameter, the CNO conductance $g_{CNO_T}$.

The membrane potential $V_q$ of neuron $q$ then obeys the equation:
\begin{equation}\label{eq:Fund}
C \frac{d V_q}{d t}= -g_{L} \pare{V_q - \cE_{L}} - g_{CNO_T} \pare{V_q - \cE_{CNO_T}} - \sum_{Y, syn} \sum_{p \in Y} g^{(Y,p)}_{q} \,\pare{V_q - \cE_Y} \, + \, i_q(t),
\end{equation}
where $\sum_{Y, syn}$ is the sum over all possible types of synaptic currents shaping the membrane potential of neuron $q$ (for example the current resulting from glutamate transmitter with NMDA receptors). The sum $\sum_{p \in Y}$ stands for the contribution of presynaptic neurons $p$ connecting to post-synaptic neuron $q$ via a synapse of type $Y$. The term $i_q$ is an external current (here, the photoreceptors current that input BCs). The synaptic conductance $g^{(Y,p)}_{q}$ depends, in general, on the pre-synaptic voltage $V_p$ and additional activation or inactivation variables. Here, we assume that it depends only on the pre-synaptic voltage and take the form $g^{(Y,p)}_{q} = \lambda^{(Y,p)}_{q} \cN^{(Y)}(V_p)$ where $\lambda^{(Y,p)}_{q}$ is a constant, positive factor, and $\cN^{(Y)}$ is the piecewise linear function introduced in \eqref{eq:Rectif}.

We shift the voltages by $\cE_L$ (i.e. replacing $V_q$ by $V_q - \cE_L$, $\cE_X$ by $\cE_X-\cE_L$ and so on). We introduce the characteristic time scale of the membrane response for neuron $q$:
%
$$
\tau_q 
= \frac{C}{g_{L} \, + \,  g_{CNO_T} \, + \,  \sum_{Y, syn} \sum_{p \in Y}\lambda^{(Y,p)}_{q} \cN^{(Y)}(V_p)}
$$
%
%
$\tau_q$  depends on the pre-synaptic neurons states (via the conductances $g^{(Y,p)}_{q}$). We can write the time dependence of voltages as $V_p(t)=V_p^\ast + \delta V_p(t)$ where $V_p^\ast$ is the rest state, i.e. the voltage reached by neurons, in the absence of stimulation, given by:
\begin{equation}\label{eq:Vrest}
V_q^\ast=\frac{g_{CNO_T} \, \cE_{CNO_T} \, + \, \sum_{Y} \sum_{p \in Y} \lambda^{(Y,p)}_{q} \, \cN^{(Y)}(V_p^\ast)  \, \cE_Y}{g_{L} \, + \,  g_{CNO_T} \, + \,  \sum_{Y, syn} \sum_{p \in Y}\lambda^{(Y,p)}_{q} \cN^{(Y)}(V_p^\ast)}.
\end{equation}
This is a complex, non linear, system of equations where the rest voltages of all neurons are dependent.

The term $\delta V_p (t)$ corresponds to the time variations of the voltage. What makes the dynamics complex are precisely these time variations inducing important non linear effects. Here, however, we are investigating effects on the retinal response when applying CNO, considering that CNO acts on a time scale quite larger than the characteristic time $\tau_q$. That is, we neglect the fluctuations $\delta V_p(t)$ and assume that the characteristic time $\tau_q$ is only controlled by the rest part of the voltages $V_p^\ast$, depending on the CNO conductance.  Thus, we can replace $\cN^{(Y)}(V_p)$ by $\cN^{(Y)}(V_p^\ast)$ in the expression of $\tau_q$: 
\begin{equation}\label{eq:tau_simplified}
\tau_q = \frac{\tau_{L}}{1 + \tau_{L} \pare{ \frac{\zeta_{T,q}}{\cE_{CNO_T}} +  \sum_{Y, syn} \frac{1}{\cE_Y} \sum_{p \in Y} W^{(Y,p)}_q \cN^{Y}(V_p^\ast)}},
\end{equation}
where $\tau_{L} = \frac{C}{g_{L}}$ is the leak characteristic time.


Finally, we introduce the synaptic weights:
\begin{equation}\label{eq:Wpq}
W^{(Y,p)}_q = \frac{\lambda^{(Y,p)}_{q} \, \cE_Y}{C},
\end{equation}
the CNO polarization:
\begin{equation}\label{eq:zetaApp}
\zeta_{T,q}= \frac{g_{CNO_T} \, \cE_{CNO_T}}{C},
\end{equation}
and the forcing term
$F_q(t)= \,\frac{i_q(t)}{C}$, and we write:
\begin{equation}\label{eq:Fund_Model_Supp}
\frac{d V_q}{dt} = -\frac{1}{\tau_q} V_q\, + \, \sum_{Y} \sum_{p \in Y} W^{(Y,p)}_q \, \cN^{(Y)}(V_p) \, + \, \zeta_{T,q} \, + \, F_q(t), \quad q=1 \dots N,
\end{equation}
where we dropped $"syn"$ in $\sum_{Y, syn}$ for legibility. This is the general form of cells voltage equations used in the paper, leading to eq. \eqref{eq:Diff_Syst}.

\sSec{Rest states, characteristic times and CNO dependence}{RestState}

We study now the dependence of the rest states and characteristic times in CNO, for the model \eqref{eq:Diff_Syst}. The details of the computations can be found on the web page \url{https://team.inria.fr/biovision/mathematicalderivationscno/}. From \eqref{eq:tau_simplified} the characteristic times take the form:
\begin{equation}\label{eq:CharacteristicTimes}
\left\{
	\begin{array}{lll}
	\tau_{B_i} &=& \frac{\tau_{L}}{1 +  \frac{\tau_{L}}{\cE_A} \sum_{j =1}^N \W{A}{j}{B}{i} \cN^{A}(V_{A_j}^\ast)},\\
	\tau_{A_j} &=& \frac{\tau_{L}}{1 + \tau_{L} \pare{ \frac{\zeta_A}{\cE_{CNO_A}} +  \frac{1}{\cE_B} \sum_{i =1}^N \W{B}{i}{A}{j} \cN^{B}(V_{B_i}^\ast)}},\\
\tau_{G_k} &=& \frac{\tau_{L}}{1 + \tau_{L} \pare{ \frac{\zeta_G}{\cE_{CNO_G}} + \frac{1}{\cE_B}  \sum_{i=1}^N \W{B}{i}{G}{k} \cN^{B}(V_{B_i}^\ast) + \frac{1}{\cE_A}  \sum_{j=1}^N \W{A}{j}{G}{k} \cN^{A}(V_{A_j}^\ast)}}.
\end{array}
\right.	
\end{equation}
Here, we see an evident and expected dependence on CNO. Increasing
$\zeta_A$ decreases the characteristic times of the membrane. This is, however, not complete, as the rest states depend themselves on CNO.
Indeed, assume, for simplicity, that the system is uniform in space, so that the characteristic times do not depend on the cell index (only on the cell type). Then, the rest states are  given by (in vector form):
\begin{equation}\label{eq:RestStateGen}
\left\{
\begin{array}{lll}
\vV^\ast_B &=& \tau_B \, \pare{\cI_N \,- \, \tau_A \tau_B \W{A}{}{B}{}  \W{B}{}{A}{}}^{-1}.\pare{\tau_A \W{A}{}{B}{} \zeta_A \, -  \, \tau_A \theta_B \, \W{A}{}{B}{} \W{B}{}{A}{} \,-\,   \theta_A  \, \W{A}{}{B}{}}\vun,\\
\vV^\ast_A &=& \tau_A \, \pare{\cI_N \,- \, \tau_A \tau_B  \W{B}{}{A}{} \W{A}{}{B}{}}^{-1}.\pare{  \zeta_A  \,-\, \tau_B  \theta_A \W{B}{}{A}{}\W{A}{}{B}{} \, -\, \theta_B \, \W{B}{}{A}{}}\vun,\\
\vV^\ast_G &=& \tau_G \W{A}{}{G}{} \vV^\ast_A \, + \, \tau_G \W{B}{}{G}{} \, \vV^\ast_B \, + \, \tau_G \pare{\zeta_G \,-\, \theta_A \W{A}{}{G}{} \,-\, \theta_B \W{B}{}{G}{}}.\vun,
\end{array}.
\right.
\end{equation}
where $\cI_N$ is the $N\times N$ identity matrix and $\vun$ is the unit vector in $N$ dimension. Here, we have assumed that the rest states are above threshold and that $\cI \,- \, \tau_A \tau_B \W{A}{}{B}{}  \W{B}{}{A}{}$ and $\cI \,- \, \tau_A \tau_B  \W{B}{}{A}{} \W{A}{}{B}{}$ are invertible. See the web page  \url{https://team.inria.fr/biovision/mathematicalderivationscno/} for the general case. 

The system of equations \eqref{eq:CharacteristicTimes} and  \eqref{eq:RestStateGen} are thus non linearly entangled. Especially, there is a  nice feedback effect appearing in  feedback "loop" terms like  $\W{A}{}{B}{}  \W{B}{}{A}{}$. Actually, e.g. the term $\pare{\cI_N \,- \, \tau_A \tau_B \W{A}{}{B}{}  \W{B}{}{A}{}}^{-1}$ involves loops  $\pare{\W{A}{}{B}{}  \W{B}{}{A}{}}^n$ of order $n>0$, with an amplitude decaying with $n$. Consider now the role of the parameter $\zeta_A$.
It impacts directly ACs and indirectly, through the network, the BCs and RGCs activity. One observes, first, a direct effect on the ACs rest state. Looking at the numerator of \eqref{eq:RestStateGen}, increasing $\zeta_A$ would have the effect of depolarizing the cell (if $\cE_{CNO_A} >0$) or hyperpolarizing it  (if $\cE_{CNO_A} < 0$), if $\tau_A$ and $\tau_B$ were independent of CNO. 
However,  CNO acts as well on these times. The net effect of $\zeta_A$ is thus non linear and cannot be anticipated by simple arguments. 

In the case of section \ref{Sec:ParametersReduction} with a reduced set of parameters the equations for characteristic times reduce to \eqref{eq:CharTimesSimple} while the rest states are given by \eqref{eq:RestStatesSimple}.

\sSec{Mathematical analysis of network dynamics}{SupMaths}

In the following, the notation $\diag\pare{x_n}_{n=1}^N$ denotes a diagonal $N\times N$ matrix with diagonal entries $x_n$. 

\ssSec{Joint dynamics}{JointDynamics}

The joint dynamics of all cells type
is given by the dynamical system \eqref{eq:Diff_Syst}.
%
%
%
%
We use Greek indices $\alpha,\beta,\gamma = 1 \dots 3N$ and define the state vector $\vcX$ with entries:  
$$\cX_\alpha =
\left\{
\begin{array}{llll}
&\V{B}{i}, \quad &\alpha=i, &i=1 \dots N; \\
&\V{A}{j}, \quad &\alpha=N+j, &j=1 \dots N; \\
&\V{G}{k}, \quad &\alpha=2N+k, &k=1 \dots N.
\end{array}
\right.
$$
We introduce $\vcF$ with entries:
$$\cF_\alpha =
\left\{
\begin{array}{llll}
&\F{B}{i}, \quad &\alpha=i, &i=1 \dots N; \\
&\zeta_A, \quad &\alpha=N+j, &j=1 \dots N; \\
&\zeta_G, \quad &\alpha=2N+k, &k=1 \dots N;
\end{array}
\right.
$$
and the rectification vector $\vcR(\vcX)$ with entries:
$$\cR_\alpha(\vcX) =
\left\{
\begin{array}{llll}
&\N{B}\pare{\V{B}{i}}, \quad &\alpha=i, &i=1 \dots N; \\
&\N{A}\pare{\V{A}{j}}, \quad &\alpha=N+j, &j=1 \dots N; \\
&0, \quad &\alpha=2N+k, &k=1 \dots N;
\end{array}
\right.
$$

We introduce the $3N \times 3N$ matrices:
\begin{equation}\label{eq:Tau}
\cT=
\pare{\begin{array}{cccccc}
& \diag\bra{\tau_{B_i}}_{i=1 \dots N}  && 0_{N N} &&0_{N N} \\
&0_{N N} & & \diag\bra{\tau_{A_j}}_{j=1 \dots N} && 0_{N N}\\
& 0_{N N}  && 0_{N N} &&\diag\bra{\tau_{G_k}}_{k=1 \dots N}
\end{array}
},
\end{equation}
characterizing the characteristic integration times of cells,
\begin{equation}\label{eq:Lc}
\cW=
\pare{\begin{array}{cccccc}
& 0_{N N}  && \W{A}{}{B}{} &&0_{N N} \\
&\W{B}{}{A}{} && 0_{N N}  && 0_{N N}\\
& \W{B}{}{G}{}  && \W{A}{}{G}{} &&0_{N N} 
\end{array}
},
\end{equation}
summarizing chemical synapses interactions. %
Then, the dynamical system \eqref{eq:Diff_Syst} reads, in vector form:
\begin{equation}\label{eq:Diff_Syst_Vect}
\frac{d \vcX}{dt} \,=\, -\, \cTm.\vcX + \cW.\vcR(\vcX) + \vcF(t).
\end{equation}

We remark that eq. \eqref{eq:Diff_Syst_Vect} has a specific product structure: the dynamics of RGCs is driven by BCs and ACs with no feedback. This means that one can study first the coupled dynamics of  BCs and ACs and then the effect on RGCs.

\ssSec{Linear evolution}{Linear_evolution}


We consider the evolution of eq. \eqref{eq:Diff_Syst_Vect} from an initial time $t_0$. Typically, $t_0$ is a reference time where the network is at rest, before the stimulus is applied. 
The dynamical system has almost the form of a non-autonomous linear system driven by the term $\vcF(t)$. There is however a weak non linearity, due to the piecewise linear rectification appearing in the term $\vcR(\vcX)$. 
Therefore, when the voltages of all cells are large enough the system is linear. Mathematically, there is a domain  of $\setR^{3N}$:
\begin{equation}\label{eq:Omega}
\cD=\Set{\V{B}{i} \geq \theta_B, \V{A}{j} \geq \theta_A, i,j=1 \dots N},
\end{equation}
where $R\pare{\vcX}$ is linear so that eq. \eqref{eq:Diff_Syst_Vect} is linear too (check \citep{cessac:22} for more details). 

From now on we consider this linear case.  
We write 
$\cL\,=\,-\, \cTm \,+\, \cW$ so that:
\begin{equation}\label{eq:L}
\cL=
\pare{\begin{array}{cccccc}
& -\diag\bra{\frac{1}{\tau_{B_i}}}_{i=1 \dots N}  && \W{A}{}{B}{} &&0_{N N} \\
&\W{B}{}{A}{} & & -\diag\bra{\frac{1}{\tau_{A_j}}}_{j=1 \dots N} && 0_{N N}\\
& \W{B}{}{G}{}  && \W{A}{}{G}{} &&-\diag\bra{\frac{1}{\tau_{G_k}}}_{k=1 \dots N}
\end{array}
},
\end{equation}
We introduce the $N$ dimensional vector $\vun=\vec(1)_{i=1}^N$, 
and the 
$3 \, N$ dimensional vector 
$
\vcC=\vect{
-\theta_A \, \W{A}{}{B}{}.\vun\\
-\theta_B \, \W{B}{}{A}{}.\vun\\
-\pare{\theta_B \, \W{B}{}{G}{}.\vun \, + \, \theta_A \, \W{A}{}{G}{}}.\vun
}$
and \eqref{eq:Diff_Syst_Vect} reads $\frac{d \vcX}{dt} =\cL.\vcX \,+\, \vcF(t) \, + \, \vcC$.\\

We assume that $\cL$ is invertible. This assumption, and more generally, the spectrum of $\cL$ is further discussed in section \ref{Sec:SpectrumL}. The general solution of eq. \eqref{eq:Diff_Syst_Vect} is:
\begin{equation}\label{eq:GenSolSDLin}
\vcX(t) = e^{\cL(t-t_0)}.\vcX(t_0) \,+\, \int_{t_0}^t e^{\cL(t-s)} \, \vcF(s) \, ds \,-\, \cL^{-1}.\bra{\cI_{3N,3N} \,-\, e^{\cL(t-t_0)}}.\vcC.
\end{equation}
where $\cI_{3N,3N}$ is the $3N$ dimensional identity matrix.

Although this equation is general, it actually stands when one can define a notion of asymptotic regime. That is, when $\cL$ has stable eigenvalues (eigenvalues with a strictly negative real part). The spectrum of $\cL$ is studied below and conditions ensuring the stability of eigenvalues are given. Here, we are going to assume that eigenvalues are all stable and that $t-t_0$ is large so that we can remove the transient term $e^{\cL(t-t_0)}.\vcX(t_0)$ depending on the initial condition $\vcX(t_0)$. In addition, the last term converges to:
\begin{equation}\label{eq:Xrest}
\vcX^\ast = -\cL^{-1}.\vcC,
\end{equation}
the rest state of the linear system, which vanishes whenever the thresholds $\theta_A,\theta_B$ are set to $0$.
\ssSec{Derivation of eq. \eqref{eq:XalphaDriven} }{Derivation_of_X}
We note the eigenvalues of $\cL$, $\lambda_\beta, \beta=1 \dots 3N$ and its eigenvectors, $\cP_\beta$ (the columns of the matrix $\cP$ transforming $\cL$ in diagonal form). We consider first the case $\vcC=\vz$.
We have then, from \eqref{eq:GenSolSDLin}:
$$
\cX_\alpha(t) = \sum_{\beta=1}^{3N}\sum_{\gamma=1}^{3N} \cP_{\alpha\beta} \cP^{-1}_{\beta \gamma} \int_{-\infty}^t e^{\lambda_\beta(t-s)} \, \cF_\gamma(s) \, ds,
$$
where $\cF_\gamma=\F{B}{i}$, $\gamma=i=1 \dots N$ (BCs).

We recall that, from  \eqref{eq:FBip}, $\F{B}{\gamma}(t)= \frac{\Vdr{drive}{\gamma}{}}{\t{}{B_\gamma}} \, + \, \frac{d \Vdr{drive}{\gamma}{}}{d t}$, so that:
%
%
$$
\int_{-\infty}^t e^{\lambda_\beta(t-s)} \, \cF_\gamma(s) \, ds
=\Vdr{drive}{\gamma}{}(t) \, + \, \CRep{\beta}{\gamma} \, \int_{-\infty}^t e^{\lambda_\beta(t-s)} \, \Vdr{drive}{\gamma}{}(s)\, ds, \quad \gamma=1 \dots N,
$$
for B cells, with $\CRep{\beta}{\gamma}=\frac{1}{\t{}{B_\gamma}} + \lambda_\beta$ and using $\Vdr{drive}{\gamma}{}(-\infty) = 0$.

For $\gamma=N+1 \dots 2N$, $\cF_\gamma=\zeta_A$ (ACs) we have:
$$
\int_{-\infty}^t e^{\lambda_\beta(t-s)} \zeta_A \, ds 
 = -\frac{\zeta_A}{\lambda_\beta}.
 $$
Finally, for $\gamma=2N+1 \dots 3N$, $\cF_\gamma=\zeta_G$ (RGCs):
$$
\int_{-\infty}^t e^{\lambda_\beta(t-s)} \zeta_G \, ds =
-\frac{\zeta_G}{\lambda_\beta}.
$$

We remark that:
\begin{equation*}
\begin{split}
\sum_{\beta=1}^{3N}\sum_{\gamma=1}^{3N} \cP_{\alpha\beta} \cP^{-1}_{\beta \gamma}\Vdr{drive}{\gamma}{}(t)
&=\sum_{\gamma=1}^{3N} \Vdr{drive}{\gamma}{}(t) \pare{\sum_{\beta=1}^{3N} \cP_{\alpha\beta} \cP^{-1}_{\beta \gamma}} \\
&= \sum_{\gamma=1}^{3N} \Vdr{drive}{\gamma}{}(t)  \delta_{\alpha\gamma} \\
&= \Vdr{drive}{\alpha}{}(t).
\end{split}
\end{equation*}
%

It follows that:
$$
\cX_\alpha(t) = \Vdr{drive}{\alpha}{}(t) \, + \E{drive}{\alpha}{}(t) \, + \,  \E{CNO_A}{\alpha}{} \, + \,  \E{CNO_G}{\alpha}{}, \quad \alpha =1 \dots 3N.
$$
with:
$$
\E{drive}{\alpha}{}(t)=\sum_{\beta=1}^{3N}\sum_{\gamma=1}^{N} \cP_{\alpha\beta} \, \cP^{-1}_{\beta \gamma}  
\CRep{\beta}{\gamma} \, \int_{-\infty}^t e^{\lambda_\beta(t-s)} \, \Vdr{drive}{\gamma}{}(s)\, ds, \quad \alpha=1 \dots 3N,
$$
$$
\E{CNO_A}{\alpha}{}= \,-\, \zeta_A   \, \sum_{\beta=1}^{3N}\sum_{\gamma=N+1}^{2N}   \frac{\cP_{\alpha\beta}\, \cP^{-1}_{\beta \gamma}}{\lambda_\beta}, \alpha = N+1 \dots 2N;
$$
$$
\E{CNO_G}{\alpha}{}= \zeta_G   \, \tau_{G}, \quad  \alpha = 2N+1 \dots 3N,
$$
where the last result comes from $\lambda_\beta \,=\, -\, \frac{1}{\tau_G}$, for $\beta=2N+1 \dots 3N$ (see next section). The expression of  $\cX_\alpha(t)$ corresponds to \eqref{eq:XalphaDriven}.

When $\cC \neq \vz$, there is an additional term corresponding to the rest state \eqref{eq:Xrest}.


\ssSec{Eigenvalues and eigenvectors of $\cL$}{SpectrumL}

\paragraph{Linear case.}
We start from the eq. \eqref{eq:L} of the linear operator ruling the dynamics in the set $\cD$ defined by \eqref{eq:Omega}. We consider, as in the main text, the case where all characteristic times $\tau_{B_i}$ are equal to $\tau_B$, all characteristic times $\tau_{A_j}$ are equal to $\tau_A$ and all characteristic times $\tau_{G_k}$ are equal to $\tau_G$. Using the same notations as the main text we have:
%
$$
\cL=
\pare{\begin{array}{cccccc}
& -\frac{\cI_{NN}}{\tau_B}  && -w^-  \, \Co{A}{}{B}{} &&0_{NN} \\
& w^+\,\Co{B}{}{A}{} & & -\frac{\cI_N}{\tau_A} && 0_{NN}\\
& \Amp{B}{G} \, \Co{B}{}{G}{}  && \Amp{A}{G} \, \Co{A}{}{G}{} &&-\frac{\cI_{NN}}{\tau_G} 
\end{array},
}
$$
where $0_{NN}$ is the $N \times N$ $0$ matrix and  $I_{NN}$ the $N \times N$ $0$ identity matrix.

\paragraph{Eigenvalues and eigenvectors.}
 We consider the case where $\Co{A}{}{B}{}=\Co{B}{}{A}{}$.  We note $\kappa_n, n=1 \dots N$, the eigenvalues of $\Co{A}{}{B}{}$ ordered as $\abs{\kappa_1} \leq \abs{\kappa_2} \leq \dots \leq \abs{\kappa_n}$ and note the normalized eigenvectors $\vpsi_n$, $n=1 \dots N$.

We seek the eigenvalues, $\lambda_\beta$, and eigenvectors, $\vcP_\beta$, $\beta=1 \dots 3N$, of $\cL$. It is evident, from the form of $\cL$, that there are $N$ eigenvalues $\lambda_\beta =-\frac{1}{\tau_G}, \vcP_\beta=\ve_\beta$ where $\ve_\beta$ is the canonical basis vector in direction $\beta$. We attribute them the indices $\beta=2N+1 \dots 3N$ as this indexing corresponds to the form of $\cL$ when $w^-=w^+=0$. We seek the $2N$ remaining eigenvalues-eigenvectors assuming that $\vcP_\beta$s has the form:
\begin{equation}\label{eq:Ansatz}
 \vcP_\beta=\vect{\vpsi_n \\ \rho_n \vpsi_n \\ \vphi_n}, \quad n=1 \dots N,
\end{equation}
where $\rho_n$ is an unknown parameter and $\vphi_n$ a $N$ dimensional vector, to be determined, from the characteristic equation:
$$
\cL.\vcP_\beta=\lambda_\beta \vcP_\beta.  
$$ 
This leads to the system of equations:
\begin{equation}\label{eq:lambda_unif}
\left\{
\begin{array}{llll}
\lambda_\beta &=& -\frac{1}{\tau_B} - w^- \, \rho_n \, \kappa_n;\\
w^+ \kappa_n - \frac{\rho_n}{\tau_A} &=&\rho_n \, \lambda_\beta;\\
\lambda_\beta \, \vphi_n&=&\pare{\Amp{B}{G} \, \Co{B}{}{G}{}  + \Amp{A}{G} \, \rho_n \, \Co{A}{}{G}{}} \, \vpsi_n -\frac{1}{\tau_G} \, \vphi_n,
\end{array},
\right.
\end{equation}
We first assume that $w^-,w^+>0$ and later discuss the limit when these quantities tend to zero. 
Combining the two first equations leads to a second-order polynomial in the variable $\rho_n$,
\begin{equation}\label{eq:Polyrho}
w^- \, \kappa_n \, \rho_n^2 \, - \, \frac{1}{\tau} \, \rho_n \, + \, w^+ \, \kappa_n=0,
\end{equation}
 giving $2$ solutions for each $n$:
\begin{equation}\label{eq:rhon}
\rho_n^\pm= 
\left\{
\begin{array}{lll}
\frac{1}{2 \, \tau \, w^- \, \kappa_n }\pare{1 \, \pm \, \sqrt{1- 4 \, \mu\, \kappa_n^2 }},& \quad \kappa_n \neq 0, \frac{1}{\tau} \neq 0;\\
w^+ \, \tau \, \kappa_n, & \quad \kappa_n =0,\frac{1}{\tau} \neq 0;\\
\pm \sqrt{- \frac{w^+}{w^-}},& \quad \frac{1}{\tau}=0.
\end{array}
\right.
\end{equation}
where:
\begin{equation}\label{eq:tau}
\frac{1}{\tau}=\frac{1}{\tau_A} - \frac{1}{\tau_B}.
\end{equation}
and:
\begin{equation}\label{eq:mu}
\mu= w^- \, w^+ \, \tau^2 \geq 0.
\end{equation}

The $2N$ first eigenvalues of $\cL$ are therefore given by:

\begin{equation}\label{eq:lambdan}
\lambda_n^\pm = 
\left\{
\begin{array}{llll}
-\frac{1}{2 \, \tau_{AB}} \mp \frac{1}{2 \, \tau} \, \sqrt{1- 4 \, \mu\, \kappa_n^2},& \quad \frac{1}{\tau} \neq 0;\\
&& \\
-\frac{1}{\tau_A} \, \mp \, \sqrt{-w^- \, w^+ \, \kappa_n^2}, 
& \quad \frac{1}{\tau} = 0.
\end{array}
\right.
\end{equation}
where:
\begin{equation}\label{eq:tauAB}
\frac{1}{\tau_{AB}}=\frac{1}{\tau_A} + \frac{1}{\tau_B}.
\end{equation}
We finally obtain $2N$ vectors $\vphi_n$:
\begin{equation}\label{eq:phin}
\vphi_n^\pm=\frac{1}{\lambda_n^\pm + \frac{1}{\tau_G}} \, \pare{\Amp{B}{G} \, \Co{B}{}{G}{}  + \Amp{A}{G} \, \rho_n^\pm \, \Co{A}{}{G}{}} \, \vpsi_n.
\end{equation}

Let us now discuss the limit when $w^-$ or $w^+$ or both tend to $0$.
If $w^- = 0$, $\rho_n=w^+ \kappa_n \tau$ from \eqref{eq:Polyrho}. If $w^+ = 0$
there are two solutions of $\eqref{eq:Polyrho}$, $\rho_n=0$ or $\rho_n=\frac{1}{\tau w^- \kappa_n}$. Finally, when $w^-=w^+=0$, $\rho_n=0$ and the ansatz \eqref{eq:Ansatz} does not apply. Actually, in this case, $\cL$  is diagonal, the $N$ first eigenvalues are $-\frac{1}{\tau_B}$, the $N$ next eigenvalues are $-\frac{1}{\tau_A}$. We have, in this case: $\lambda_n^+ = -\frac{1}{\tau_B}  $ and $\lambda_n^- = -\frac{1}{\tau_A}$. Therefore, we order eigenvalues and eigenvectors of $\cM$ such that the $N$ first eigenvalues are $\lambda_n^+, n=1 \dots N$, and the $N$ next are  $\lambda_n^-, n=N+1 \dots 2N.$

We finally end up with the following form for the eigenvalues and eigenvectors of $\cL$:

\begin{equation}\label{eq:EigenvectorsL}
\resizebox{\textwidth}{!}{$
\begin{array}{lllll}
\lambda_\beta = \lambda_n^+, \vcP_\beta=\vect{\vpsi_n \\ \rho^+_n \vpsi_n \\ \frac{1}{\lambda_n^+ + \frac{1}{\tau_G}} \, \pare{\Amp{B}{G} \, \Co{B}{}{G}{}  + \Amp{A}{G} \, \rho_n^+ \, \Co{A}{}{G}{}} \, \vpsi_n}, \quad& \beta=n=1 \dots N,&\\
\lambda_\beta =\lambda_n^-,  \vcP_\beta=\vect{\vpsi_n \\ \rho^-_n \vpsi_n \\ \frac{1}{\lambda_n^- + \frac{1}{\tau_G}} \, \pare{\Amp{B}{G} \, \Co{B}{}{G}{}  + \Amp{A}{G} \, \rho_n^- \, \Co{A}{}{G}{}} \, \vpsi_n}, &\beta=N+1 \dots 2N, &n=1 \dots N,\\
\lambda_\beta =-\frac{1}{\tau_G}, \vcP_\beta=\ve_\beta, &\beta=2N+1 \dots 3N.
\end{array}$}
\end{equation}

\paragraph{Skeleton.}
The eigenvalues $\lambda_n^\pm$ in \ref{eq:lambdan} can be real or complex conjugated.
By increasing $\mu$, they become complex when:
\begin{equation}\label{eq:mucomplex}
\mu > \frac{1}{4 \, \kappa_n} \equiv \mu_{n,c}.
\end{equation}
In this case the real part is $-\frac{1}{2 \, \tau_{AB}}$, the imaginary part is $\pm \frac{1}{2 \, \tau} \, \sqrt{1- 4 \, \mu\, \kappa_n^2 }$. If $\mu \leq \mu_{n,c}$, eigenvalues $\lambda_\beta$ are real with a negative real part. This implies that the linear dynamical system \eqref{eq:Diff_Syst_Vect} is stable. 

The N equations \eqref{eq:mucomplex} define what we call the "skeleton of the RFs map". In the main text, we introduced  the quantities $r=\frac{\tau_A}{\tau_B},s=\frac{w^-}{w^+}$.
In these variables, the critical condition \eqref{eq:mucomplex} reads:
\begin{equation}\label{eq:skeleton}
s_{n_c}=\frac{1}{w^{+^2} \, \tau_B^2} \, \frac{1}{4 \,\kappa_n^2} \, \frac{\pare{1-r}^2}{r^2}.
\end{equation}
This defines two critical lines symmetric with respect to $r=1$. These lines are invariant by the variable change $r'=r,s'=\pare{\frac{{w'}^+ \, {\tau'}_B}{w^+ \, \tau_B} }^2 \, s$. This allows to map the skeleton obtained from a set of values ${\tau'}_B,{w'}^+$ to the skeleton
obtained with references value $\tau_B,w^+$. 

\paragraph{Rectification.} In this paper we have essentially considered a situation where cells are not rectified, whereas the full model (eq. \eqref{eq:Diff_Syst}) considers rectification, in agreement with realistic biological systems. The mathematical effect of rectification of a cell $\Cell{A}{j}$, is to set to zero the corresponding row in the matrix $\W{A}{}{B}{}$. This has several consequences. First, we cannot apply anymore the useful Ansatz used in the section, that is $\Co{A}{}{B}{}= \Co{B}{}{A}{}$. In addition, the vanishing of only one row in $\cL$ completely modifies its spectrum. However, thresholding in rectification corresponds to partition the phase space of the model, a compact subset of $\setR^{3N}$,  into convex subdomains delimited by hyperplanes. In each of these domains the matrix   $\cW.\vcR(\vcX)$ appearing in eq. \eqref{eq:Diff_Syst_Vect} is linear with a number of zero eigenvalues corresponding to the number of rectified cells. This matrix acts as a projector on the complementary subspace of its kernel. In each of these subdomains eq.  \eqref{eq:XalphaDriven} applies. One can actually compute, for a given stimulus, the time of entrance and exit in a new subdomain with the effect of modifying the eigenvalues and eigenvectors appearing in eq.  \eqref{eq:XalphaDriven}. The resulting equation is quite complex though and will require further investigations.  See \citep{cessac:20} for more details.

\ssSec{Nearest neighbours connectivity}{Nearest_neighbours}

We  consider the case where the connectivity matrices $\Co{B}{}{A}{}=\Co{A}{}{B}{}$ have nearest neighbours symmetric connections. 
We also assume that the dynamics hold on a square lattice with 
null boundary conditions. 
We define $\alpha=i_x \in \Set{1 \dots L=N}$ in one dimension and $\alpha=i_x + (i_y-1) \, L \in \Set{1 \dots L^2=N}$ in two dimensions. We also set  $n=n_x \in \Set{1 \dots L=N}$ in one dimension and $n=n_x + (n_y-1) \, L \in \Set{1 \dots L^2=N}$ in two dimensions. 
Then, the eigenvectors and eigenvalues of these matrices have the form:
\begin{equation}\label{eq:EigenvectorsDelta}
\begin{array}{lll}
\psi_{\alpha,n} = \pare{\frac{2}{L+1}}^\frac{d}{2} \prod_{l} \, \sin\pare{\frac{n_l \pi}{L+1} \, i_l},\\
\kappa_n = 2 \, \sum_l \bra{\cos\pare{\frac{n_l \pi}{L+1}}}; 
\end{array}
\end{equation}
with $l=x$ for $d=1$ and $l=x,y$ for $d=2$. Especially, in one dimension:
$$
\psi_{\alpha,n} = \sqrt{\frac{2}{L+1}} \sin\pare{\frac{\alpha \pi}{L+1} \, n},
$$

The quantum numbers $\pare{n_x,n_y}$ define a wave vector $\vk_n=\pare{\frac{n_x \pi}{L+1},\frac{n_y \pi}{L+1}}$ corresponding to wave lengths $\pare{\frac{L+1}{ n_x},\frac{L+1}{ n_y}}$. 
Hence, the first eigenmode $(n_x=1,n_y=1)$ corresponds to the largest space scale (scale of the whole retina) with the smallest eigenvalue (in absolute value) $s_{(1,1)}=2\pare{\cos\pare{\frac{\pi}{L+1}}+\cos\pare{\frac{\pi}{L+1}}-2}$. Each of these eigenmodes is related to a characteristic time $\tau_n=\frac{1}{\lambda_n}$.

Eigenvalues $\kappa_n$ can be positive or negative. However, from eq. \eqref{eq:lambdan} this has no impact on the eigenvalues as what matters is $\kappa_n^2$. This induces however a symmetry $\kappa_n \to -\kappa_n$ that can be seen in the skeleton Figure \ref{Fig:PhaseMap}, not forgetting that this figure is in log scale.

\bibliographystyle{apalike}
\bibliography{odyssee,References}

\end{document}